\documentclass[prd,aps,showpacs,notitlepage,superscriptaddress,nofootinbib]{revtex4-1}

\usepackage{amsmath}
\usepackage{multirow}
\usepackage{float}
\usepackage{soul}
\usepackage{caption}
\usepackage {tikz}

\def\bea#1\eea{\begin{align}#1\end{align}}

\makeatletter
\def\slash#1{{\mathpalette\c@ncel{#1}}} 
\makeatother

\newcommand\beq{\begin{eqnarray}}
\newcommand\eeq{\end{eqnarray}}

\newcommand\la{\langle}
\newcommand\ra{\rangle}

\begin{document}
\title{Extension of the non-pole technique to the twist-3 gluon distribution contribution in $pp$ collisions}

\date{\today}

\author{Longjie Chen}
\email{longjie.chen@ifj.edu.pl}
\affiliation{Institute of Nuclear Physics Polish Academy of Sciences,
ul. Radzikowskiego 152, PL-31-342 Krak\'ow, Poland}

\author{Shinsuke Yoshida}
\email{shinyoshida85@gmail.com}
\affiliation{State Key Laboratory of Nuclear Physics and Technology, Institute of
Quantum Matter, South China Normal University, Guangzhou 510006, China}
\affiliation{Guangdong Basic Research Center of Excellence for
Structure and Fundamental Interactions of Matter, Guangdong
Provincial Key Laboratory of Nuclear Science, Guangzhou
510006, China}

\begin{abstract}

It is known that the Sivers effect is described as a twist-3 effect within the collinear factorization framework and has the characteristic feature that it arises from the pole contribution in a hard parton scattering. The calculation formalism that focuses only on the pole part was established in the 2000s and 
SSAs were calculated for many processes in $ep$ and $pp$ collisions.
On the other hand, there also exists a contribution from twist-3 fragmentation functions regarded as 
the Collins effect. The Collins effect arises from the nonpole part of the hard scattering and its calculation is formulated in a somewhat different manner from that for the Sivers effect. In recent years, some attempts have been made to revisit the Sivers effect by applying the ``nonpole'' formalism developed for the Collins effect. For $ep$ collisions, this approach has successfully reproduced the known results. For $pp$ collisions, however, the presence of both initial-state-interaction and final-state-interaction makes the calculation more complicated and the known results have not yet been reproduced.
In this paper, we provide a solution to this problem and develop the nonpole method so that it can be applied to the Sivers effect in any process.

\end{abstract}

\maketitle


\section{Introduction}

Study of the Sivers effect through measurements of single transverse-spin asymmetries(SSAs) 
has attracted considerable attention because it provides new insights beyond the conventional 
picture of the nucleon structure, including its three-dimensional structure and 
the correlation between partonic orbital motion and the nucleon's spin.
Understanding the Sivers effect is one of the major physics goals of the next-generation Electron-Ion 
Collider(EIC)\cite{Accardi:2012qut}. 
A distinctive feature of the EIC is that it will cover not only the region of small
transverse momentum of a final-state particle, where the TMD framework\cite{Boussarie:2023izj} 
is applicable, but also the 
large transverse momentum region that was not covered in previous electron-proton collision experiments. 
In the latter region, analyses of the SSAs based on the twist-3 mechanism within collinear factorization 
are also required.
The quark Sivers functions have already been well constrained through global fits to existing SSA data
\cite{Anselmino:2005ea,Anselmino:2012aa,Bury:2020vhj,Bury:2021sue,Fernando:2023obn}. 
Furthermore, analyses that take into account relations between TMD functions and corresponding 
collinear functions have successfully provided a systematic description of the SSA data over a wide range 
of the transverse momentum\cite{Cammarota:2020qcw}. 
In contrast, a limited amount of experimental data is currently available
for the gluon Sivers function and therefore its functional form remains poorly known. 
Performing similar analyses for the gluon Sivers function is an important goal of future 
experiments to obtain a deeper understanding of the orbital motion of gluons.

The formulation of the twist-3 mechanism in collinear factorization was actively developed around
the beginning of this century inspired by measurements at Relativistic Heavy Ion Collider(RHIC)
\cite{Adams:2003fx,PHENIX:2005jxc,STAR:2008ixi,BRAHMS:2008doi,PHENIX:2010hqq,STAR:2012ljf,STAR:2013zyt,PHENIX:2013wle,PHENIX:2014qwb,PHENIX:2017wbv,PHENIX:2018qvl,PHENIX:2019ouo,PHENIX:2021irw,PHENIX:2020mft,STAR:2020grs,STAR:2020nnl,PHENIX:2021dzj,PHENIX:2022znm,PHENIX:2023axd}. 
Na\"{i}ve $T$-odd observables such as the SSA require a complex phase. 
This phase is generated by the hard scattering of partons in the case of the Sivers effect. 
In the early formulations, only the imaginary delta function part in the following decomposition of an internal propagator was taken into account in the derivation of an analytic cross section formula for 
the SSA\cite{Eguchi:2006mc}.
\beq
{1\over p^2+i\epsilon}=P{1\over p^2}-i\pi\delta(p^2).
\eeq
Using this approach, SSA formulas have been derived for many processes in $ep$
and $pp$ collisions\cite{Qiu:1998ia,Kouvaris:2006zy,Eguchi:2006qz,Koike:2009ge,Kang:2008qh,Kang:2008ih,Beppu:2010qn,Koike:2011mb,Koike:2011nx,Beppu:2013uda}.
Subsequently, twist-3 fragmentation functions associated with the Collins effect, which is 
another possible source of the SSA, were formulated\cite{Metz:2012ct,Kanazawa:2013uia}.
In the case of the Collins effect, the imaginary part of a twist-3 fragmentation function provides the complex phase and the real principal value part of the above propagator is picked up in the derivation.  
In the early stages, the formulation differed depending on whether the real or imaginary part of 
the propagator is picked up, and consequently the calculation framework of the SSA was somewhat 
different between the Sivers type and the Collins type.
In particular, in the nonpole approach established through the formulation of the Collins effect, gauge invariance and frame independence of the results are not obvious. 
They have to be demonstrated by using relations among the intrinsic, kinematical, and dynamical functions based on the equations of motion and translational invariance of the relevant matrix 
elements\cite{Kanazawa:2015ajw}.

As a recent development, the Sivers effect has been reformulated using this nonpole 
approach\cite{Xing:2019ovj,Zhang:2020wad,Yoshida:2022vnf}. 
This method calculates the Sivers effect without using the decomposition of the propagator
mentioned above. The SSA is generally expressed in terms of intrinsic, kinematical, and dynamical functions 
and relations among these functions are required to demonstrate gauge invariance 
and frame independence as in the Collins case. For the Sivers effect, the intrinsic contribution 
generally cancels, while the kinematical and dynamical functions are related to each other.
This approach not only makes it possible to calculate the Sivers and Collins effects within the same framework but is also important from a practical point of view. Recent next-to-leading-order calculations have demonstrated that the nonpole approach, which takes the kinematical function into account 
and rewrites it in terms of the dynamical function, makes the calculation simpler than that in 
the old approach\cite{Xing:2019nof,Rein:2025qhe}.
For the Sivers effect, the nonpole approach has succeeded in reproducing results consistent with those obtained by the old approach in semi-inclusive deep inelastic scattering (SIDIS) for both the twist-3 quark-gluon and 3-gluon distribution functions.
In $pp$ collisions, however, the coexistence of initial-state-interaction(ISI) and final-state-interaction(FSI) makes the formulation more complicated and no results based on the nonpole approach have been obtained to date. In this paper, we provide a solution to this problem for the case of the 3-gluon distribution
associated with the gluon Sivers effect
and demonstrate that the nonpole approach gives results consistent with those obtained by the old approach 
in $pp$ collisions as well.

The remainder of this paper is organized as follows: 
In Sec. II, we will give an overview of the established formulation for SIDIS and explain its essential 
features. In particular, we will emphasize that the sign of the $i\epsilon$ prescription is crucial 
for the Sivers effect. 
In Sec. III. A, we show that the Ward-Takahashi identities(WTIs) can be decomposed into minimal sets
in $pp$ collisions and consequently the sign of $i\epsilon$ is uniquely determined.
In Sec. III. B, we will derive an analytic cross section formula for the SSA based on the minimal WTIs. 
In Sec. III. C, we will calculate the hard cross sections and confirm that the nonpole formalism 
gives results consistent with those obtained using the old approach.
Sec. IV is devoted to a summary of our study.


\section{Overview of the nonpole technique in SIDIS}

\begin{figure}[H]
\begin{center}
  \includegraphics[height=4cm,width=12cm]{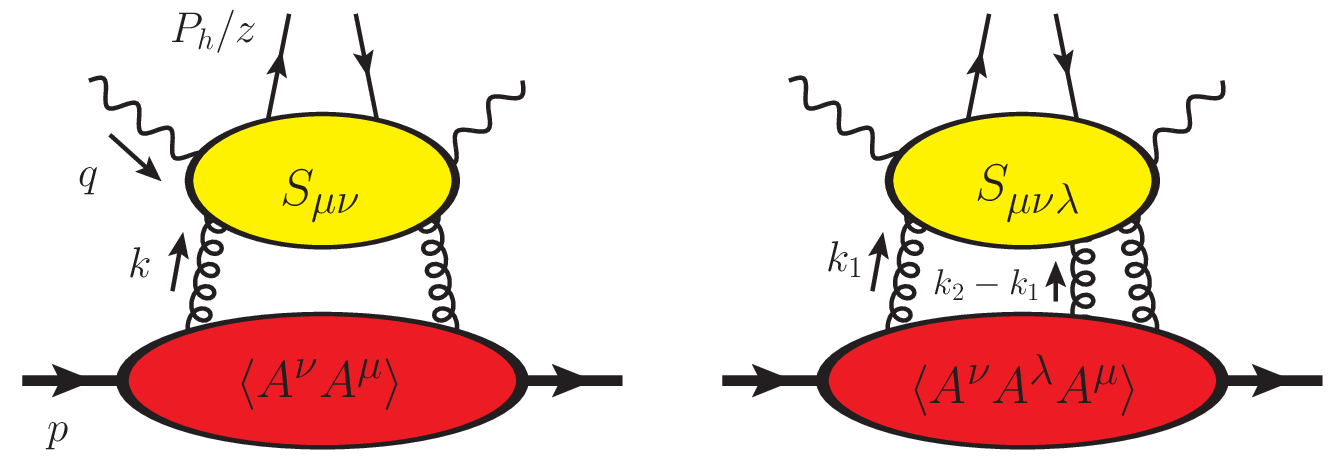}
\end{center}
 \caption{The hadronic tensor $w_{\rho\sigma}(p,q,{P_h\over z})$ in the diagrammatic method.}
\label{diagram_SIDIS}
\end{figure}

The nonpole technique for the 3-gluon distribution contribution
was developed in SIDIS\cite{Yoshida:2022vnf},
\beq
e(\ell)+p^{\uparrow}(p)\to e(\ell')+h(P_h)+X,
\eeq 
within the standard diagrammatic method\cite{Ellis:1982wd,Ellis:1982cd}.
The contribution of the 3-gluon distribution to
the polarized cross section takes the form of
\beq
\frac{d^{6}\Delta\sigma}{dx_{bj}dQ^{2}dz_{f}dP^{2}_{h}d\phi d\chi}  =
\frac{\alpha ^{2}_{em}}{128\pi^4 z_fS^{2}_{ep} x^{2}_{bj} Q^{2}}
L^{\rho\sigma}(\ell,\ell')\int{dz\over z^2}D(z)w_{\rho\sigma}(p,q,{P_h\over z}), 
\eeq
which is written in terms of the Lorentz-invariant variables
\beq
S_{ep}&=&(p+\ell)^2,\hspace{5mm}
Q^2=-q^2=-(\ell-\ell')^2,\hspace{5mm}
x_{bj}=\frac{Q^2}{2p\cdot q},\hspace{5mm}
z_f=\frac{p\cdot P_h}{p\cdot q}.
\eeq
$\alpha_{em}$ is the QED coupling, $L^{\rho\sigma}=2(\ell^{\rho}\ell^{\prime\sigma}
+\ell^{\sigma}\ell^{\prime\rho}-Q^2/2g^{\rho\sigma})$ is the leptonic tensor
and $D(z)$ is the fragmentation function of the final state hadron $h$.
The hadronic tensor $w_{\rho\sigma}$ describes the interaction between
the proton and the virtual photon emitted by the electron.
We need to consider only the two types of diagrams shown in Fig. \ref{diagram_SIDIS}
at twist-3 accuracy.
\noindent
The sum of these diagrams gives the following mathematical expression.
\beq
w(p,q,{P_h\over z})&=&w^{\rm Fig.1left}(p,q,{P_h\over z})+w^{\rm Fig.1right}(p,q,{P_h\over z}),
\\
w^{\rm Fig.1left}(p,q,{P_h\over z})&=&\int d^4\xi\int{d^4k\over (2\pi)^4}\,e^{ik\cdot \xi}\la pS|
A^{\nu}_b(0)A^{\mu}_a(\xi)|pS\ra S^{ab}_{\mu\nu}(k),
\\
w^{\rm Fig.1right}(p,q,{P_h\over z})&=&g\int d^4\xi\int d^4\eta\int{d^4k_1\over (2\pi)^4}
\int{d^4k_2\over (2\pi)^4}\,e^{ik_1\cdot \xi}e^{i(k_2-k_1)\cdot \eta}
\la pS|A^{\nu}_b(0)A^{\lambda}_c(\eta)A^{\mu}_a(\xi)|pS\ra S^{abc}_{\mu\nu\lambda}(k_1,k_2),
\label{hadronic}
\eeq
where the Lorentz indices $\rho$ and $\sigma$ of the external virtual photon lines are omitted for simplicity.
The WTIs for the hard parts $S^{ab}_{\mu\nu}(k)$
and $S^{abc}_{\mu\nu\lambda}(k_1,k_2)$ are given by
\beq
k^{\mu}S_{\mu\nu}(k)&=&k^{\nu}S_{\mu\nu}(k)=0,
\label{WTI1}
\\
(k_2-k_1)^{\lambda}S^{abc}_{\mu\nu\lambda}(k_1,k_2)&=&if^{abc}\Bigl(S_{\mu\nu}(k_2)
-S_{\mu\nu}(k_1)\Bigr),
\label{WTI2}
\\
k_1^{\mu}S^{abc}_{\mu\nu\lambda}(k_1,k_2)&=&-if^{abc}S_{\lambda\nu}(k_2),
\label{WTI3}
\\
k_2^{\nu}S^{abc}_{\mu\nu\lambda}(k_1,k_2)&=&if^{abc}S_{\mu\lambda}(k_1),
\label{WTI4}
\eeq
where $S_{\mu\nu}(k)=\delta^{ab}S_{\mu\nu}^{ab}(k)/(N_c^2-1)$. 
We decompose the momentum vectors as
\beq
k_i^{\mu}=(k_i\cdot n)p^{\mu}+\omega^{\mu}_{\ \alpha}k_i^{\alpha},
\eeq
where $\omega^{\mu}_{\ \alpha}=g^{\mu}_{\ \alpha}-p^{\mu}n_{\alpha}$
and $n$ is an arbitrary lightlike vector which satisfies $n^2=0$ and $p\cdot n=1$.
The following identities, derived from the above WTIs, are necessary for deriving the twist-3 
cross section formula.
\beq
p^{\mu}S_{\mu\nu}(k)&=&-{1\over k\cdot n}\omega^{\mu}_{\ \alpha}k^{\alpha}S_{\mu\nu}(k),
\hspace{5mm}
p^{\nu}S_{\mu\nu}(k)=-{1\over k\cdot n}\omega^{\nu}_{\ \beta}k^{\beta}S_{\mu\nu}(k),
\label{Ward1}
\\
p^{\lambda}S^{abc}_{\mu\nu\lambda}(k_1,k_2)&=&{1\over (k_2-k_1)\cdot n-i\epsilon}
\Bigl[-\omega^{\lambda}_{\ \gamma}(k_2-k_1)^{\gamma}S^{abc}_{\mu\nu\lambda}(k_1,k_2)
+if^{abc}\Bigl(S_{\mu\nu}(k_2)
-S_{\mu\nu}(k_1)\Bigr)
\Bigr],
\label{SIDIS_Ward1}
\\
p^{\mu}S^{abc}_{\mu\nu\lambda}(k_1,k_2)&=&{1\over k_1\cdot n-i\epsilon}
\Bigl[-\omega^{\mu}_{\ \alpha}k_1^{\alpha}S^{abc}_{\mu\nu\lambda}(k_1,k_2)
-if^{abc}S_{\lambda\nu}(k_2)
\Bigr],
\label{SIDIS_Ward2}
\\
p^{\nu}S^{abc}_{\mu\nu\lambda}(k_1,k_2)&=&{1\over k_2\cdot n+i\epsilon}
\Bigl[-\omega^{\nu}_{\ \beta}k_2^{\beta}S^{abc}_{\mu\nu\lambda}(k_1,k_2)
+if^{abc}S_{\mu\lambda}(k_1)
\Bigr],
\label{SIDIS_Ward3}
\eeq
Note that the WTIs (\ref{WTI2})-(\ref{WTI4}) are process-independent relations, 
whereas (\ref{SIDIS_Ward1})-(\ref{SIDIS_Ward3}), which are derived from those WTIs, 
are process-dependent due to an ambiguity in the sign of the $i\epsilon$ prescription. 
The signs of $i\epsilon$ are 
uniquely determined in SIDIS because of the fact that only FSI is present.
Using these relations, one can derive the twist-3 cross section formula as discussed
in the appendix of \cite{Yoshida:2022vnf},
\beq
w(p,q,{P_h\over z})&=&
\omega^{\mu}_{\ \alpha}\omega^{\nu}_{\ \beta}\int{dx\over x^2}
\Phi^{\alpha\beta}(x)S_{\mu\nu}(xp)
+\omega^{\mu}_{\ \alpha}\omega^{\nu}_{\ \beta}\omega^{\lambda}_{\ \gamma}
\int{dx\over x^2}\Phi^{[-]\alpha\beta\gamma}_{\partial}(x)
{\partial\over \partial k^{\lambda}}S_{\mu\nu}(k)\Bigr|_{k=xp}
\nonumber\\
&&-{1\over 2}\omega^{\mu}_{\ \alpha}\omega^{\nu}_{\ \beta}\omega^{\lambda}_{\ \gamma}
\int dx_1\int dx_2\,\Phi^{\alpha\beta\gamma}_{F\,abc}(x_1,x_2)
{1\over x_1-i\epsilon}{1\over x_2+i\epsilon}
{1\over x_2-x_1-i\epsilon}S^{abc}_{\mu\nu\lambda}(x_1p,x_2p).
\label{formula_SIDIS}
\eeq
Each gauge-invariant matrix element is given by
\beq
\Phi^{\alpha\beta}(x)=\int{d\lambda\over 2\pi}\,e^{i\lambda x}
\la pS|F_b^{\beta n}(0)[0,\lambda n]_{ba}F_a^{\alpha n}(\lambda n)|pS\ra,
\eeq
\beq
\Phi^{[\pm]\alpha\beta\gamma}_{\partial}(x)&=&
i\int{d\lambda\over 2\pi}\,e^{i\lambda x}\Bigl[\la pS|F_e^{\beta n}(0)[0,\lambda n]_{eb}
D_{ba}^{\gamma}(\lambda n)F_a^{\alpha n}(\lambda n)|pS\ra
\nonumber\\
&&+ig\int^{\mp\infty}_{\lambda}d\mu\,\la pS|F_{e}^{\beta n}(0)[0,\mu n]_{eb}
(if^{abc})F_c^{\gamma n}(\mu n)[\mu n,\lambda n]_{ad}F_d^{\alpha n}(\lambda n)|pS\ra\Bigr],
\eeq
\beq
\Phi^{\alpha\beta\gamma}_{F\,abc}(x_1,x_2)&=&i
\int{d\lambda\over 2\pi}\int{d\mu\over 2\pi}e^{i\lambda x_1}e^{i\mu(x_2-x_1)}
\la pS|F_{e}^{\beta n}(0)[0,\mu n]_{eb}
F_c^{\gamma n}(\mu n)[\mu n,\lambda n]_{ad}F_d^{\alpha n}(\lambda n)|pS\ra,
\eeq
where $[0,\lambda n]$ represents the Wilson line in the adjoint representation,
\beq
[0,\lambda n]_{ba}={\rm P}\exp\Bigl(ig\int^0_{\lambda} d\mu\,A^n(\mu n)\Bigr)_{ba}.
\eeq
One finds that the signs of $i\epsilon$ in the last line originate from those in the identities
(\ref{SIDIS_Ward1})-(\ref{SIDIS_Ward3}) upon changing the integration
variables $k_i\cdot n\to x_i$. These signs are crucial in the calculation of the Sivers type effect
because they determine the sign of the pole contribution,
\beq
{1\over x-x'\mp i\epsilon}-{1\over x-x'\pm i\epsilon}=\pm 2\pi i\delta(x-x').
\eeq
We expect that the pole structure becomes more complicated in $pp$ collisions because 
the ISI is additionally present. Our aim is to derive
the formulas corresponding to (\ref{SIDIS_Ward1})-(\ref{SIDIS_Ward3})
and (\ref{formula_SIDIS}) in $pp$ collisions.


\section{Calculation of the SSA in $J/\psi$ production in $pp$ collisions}

\subsection{Twist-2 unpolarized cross section in $gq\to qg$ channel}

\begin{figure}[H]
\begin{center}
  \includegraphics[height=5cm,width=13cm]{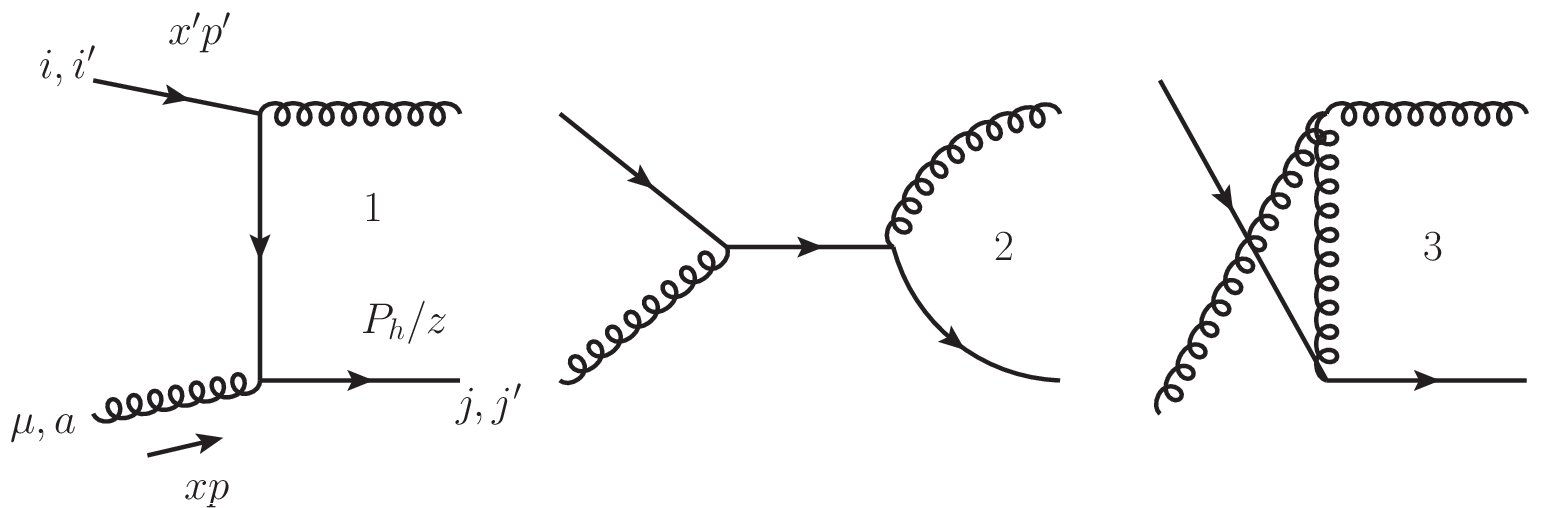}
\end{center}
 \caption{Amplitude of the quark fragmentation channel in $gq\to qg$ scattering.
 Each numbered diagram is decomposed into the color part $(C_k^a)_{j'i'}$ and the spinor part
 $({\cal M}_{k\,\mu}(xp))_{ji}$ as $(C_k^a)_{j'i'}({\cal M}_{k\,\mu}(xp))_{ji}\ (k=1,2,3)$.}
\label{unpol_amp}
\end{figure}

We consider the SSA in light hadron production in $pp$ collisions,
\beq
p^{\uparrow}(p)+p(p')\to h(P_h)+X.
\eeq
We focus on the quark fragmentation channel in the $gq\to qg$ hard parton scattering. In this case,
the unpolarized cross section is given by
\beq
P^0_{h}{d\sigma\over d^3\vec{P}_h}
={\alpha_s\over S}\int {dx\over x}G(x)\int {dx'\over x'}q(x')\int {dz\over z^2}D(z)
\Bigl(-{1\over 2}g_{\perp}^{\mu\nu}(p)\Bigr)S_{\mu\nu}(xp),
\eeq
\beq
S_{\mu\nu}(xp)=H_{\mu\nu}(xp)\delta\Bigl((xp+x'p'-{P_h\over z})^2\Bigr),
\eeq
where $\alpha_s$ is the QCD coupling, $S=(p+p')^2$ is the center of mass energy squared,
$G(x)$ and $q(x)$ are the unpolarized gluon and quark distribution functions of the proton
and $g_{\perp}^{\mu\nu}(p)=g^{\mu\nu}-p^{\mu}n^{\nu}-p^{\nu}n^{\mu}$.
The hard scattering part $H_{\mu\nu}(xp)$ is obtained by taking the square of the amplitude
shown in Fig. \ref{unpol_amp}.
\noindent
Thus the hard part $H_{\mu\nu}(xp)$ is given by
\beq
H_{\mu\nu}(xp)=\sum_{k,l=1}^{3}C_{kl}{\rm Tr}[{\cal M}_{k\,\mu}(xp)
{\cal M}_{l\,\nu}^{\dagger}(xp)].
\eeq
The color factors are evaluated as
\beq
{\rm Tr}[C^a_kC^{b\dagger}_l]=C_{kl}\delta^{ab},\hspace{5mm}
C_{kl}=\left(
  \begin{array}{ccc}
      \hspace{2mm}{C_F\over 2N_c} \hspace{2mm} & 
      \hspace{2mm}{C_F\over 2N_c}-{1\over 4}\hspace{2mm}
       & \hspace{2mm}-{1\over 4}\hspace{2mm}  \\ \\
      {C_F\over 2N_c}-{1\over 4} & {C_F\over 2N_c} & {1\over 4} \\ \\
      -{1\over 4} & {1\over 4} & {1\over 2}
    \end{array}
  \right),
\label{unpol_color}
\eeq
where $C_F=(N_c^2-1)/(2N_c)$ with $N_c=3$. One can derive the well-known unpolarized cross section
\beq
P^0_{h}{d\sigma\over d^3\vec{P}_h}
={\alpha_s\over S}\int {dx\over x}G(x)\int {dx'\over x'}q(x')\int {dz\over z^2}D(z)
\hat{\sigma}_{U}\delta(\hat{s}+\hat{t}+\hat{u}),
\eeq
where Mandelstam variables are defined as
\beq
\hat{s}=(xp+x'p')^2,\hspace{5mm}
\hat{t}=(xp-{P_h\over z})^2,\hspace{5mm}
\hat{u}=(x'p'-{P_h\over z})^2.
\eeq
The hard cross section $\hat{\sigma}_{U}$ is computed as
\beq
\hat{\sigma}_U&=&\sum_{k,l=1}^3C_{kl}\Bigl(-{1\over 2}g_{\perp\mu\nu}(p)\Bigr)
{\rm Tr}[{\cal M}^{\mu}_k(xp){\cal M}^{\nu\dagger}_l(xp)]
\nonumber\\
&=&-{C_F\over N_c}\Bigl({\hat{s}\over \hat{t}}+{\hat{t}\over \hat{s}}\Bigr)
+{\hat{s}^2+\hat{t}^2\over \hat{u}^2}.
\eeq

\subsection{Reexamination of the Ward-Takahashi identities in $pp$ collisions}

\begin{figure}[H]
\begin{center}
  \includegraphics[height=5cm,width=13cm]{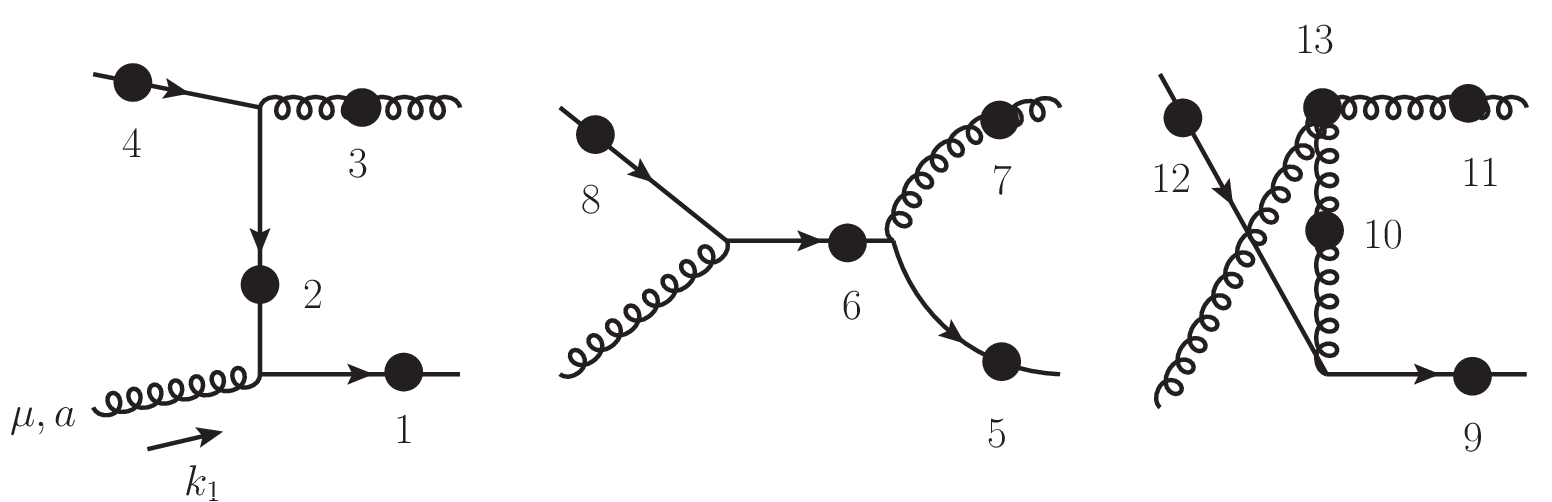}
\end{center}
 \caption{The diagrams contributing to the amplitude ${\cal M}^{ac}_{L\mu\lambda}(k_1,k_2-k_1)$.
 The gluon line with momentum $k_2-k_1$, Lorentz index $\lambda$ and color index $c$ is
 connected to one of the black dots. The diagram with the gluon line connected 
 to $k$-th black dot gives ${\cal M}_{kL\mu\lambda}$.}
\label{SL_amp}
\end{figure}

We  consider the twist-3 gluon contribution to the polarized cross section
\beq
P^0_{h}{d\Delta\sigma\over d^3\vec{P}_h}
={\alpha_s\over S}\int {dx'\over x'}q(x')\int {dz\over z^2}D(z)
\,w(p,x'p',{P_h\over z}).
\eeq
We can use the same expression as (\ref{hadronic}) for $w(p,x'p',{P_h\over z})$ just replacing
the virtual photon line with momentum $q$ with the quark line with momentum $x'p'$.
The hard part $S_{\mu\nu\lambda}(k_1,k_2)$ is given by inserting an extra gluon 
with momentum $k_2-k_1$ in each diagram in Fig. \ref{unpol_amp}.
We decompose $S_{\mu\nu\lambda}(k_1,k_2)$ into two parts,
\beq
S_{\mu\nu\lambda}^{abc}(k_1,k_2)&=&S_{L\mu\nu\lambda}^{abc}(k_1,k_2)
+S_{R\mu\nu\lambda}^{abc}(k_1,k_2),
\eeq
\beq
S_{L\mu\nu\lambda}^{abc}(k_1,k_2)&=&{\rm Tr}[{\cal M}^{ac}_{L\mu\lambda}(k_1,k_2-k_1)
(\sum_k C^b_k{\cal M}_{k\,\nu}(k_2))^\dagger]\delta\Bigl((k_2+x'p'-{P_h\over z})^2\Bigr),
\label{SL}
\\
S_{R\mu\nu\lambda}^{abc}(k_1,k_2)&=&{\rm Tr}[(\sum_k C^a_k{\cal M}_{k\,\mu}(k_1))
{\cal M}^{bc}_{R\nu\lambda}(k_2,k_2-k_1)]\delta\Bigl((k_1+x'p'-{P_h\over z})^2\Bigr).
\label{SR}
\eeq
where ${\rm Tr}[\cdots]$ denotes the trace over both the color and the spinor indices.
There are 13 diagrams in ${\cal M}^{ac}_{L\mu\lambda}(k_1,k_2-k_1)$ as shown in
Fig. \ref{SL_amp}.
\noindent
An important difference between SIDIS and $pp$ collisions is that
there are both ISI and FSI in the latter.
\begin{figure}[H]
\begin{center}
  \includegraphics[height=5cm,width=10cm]{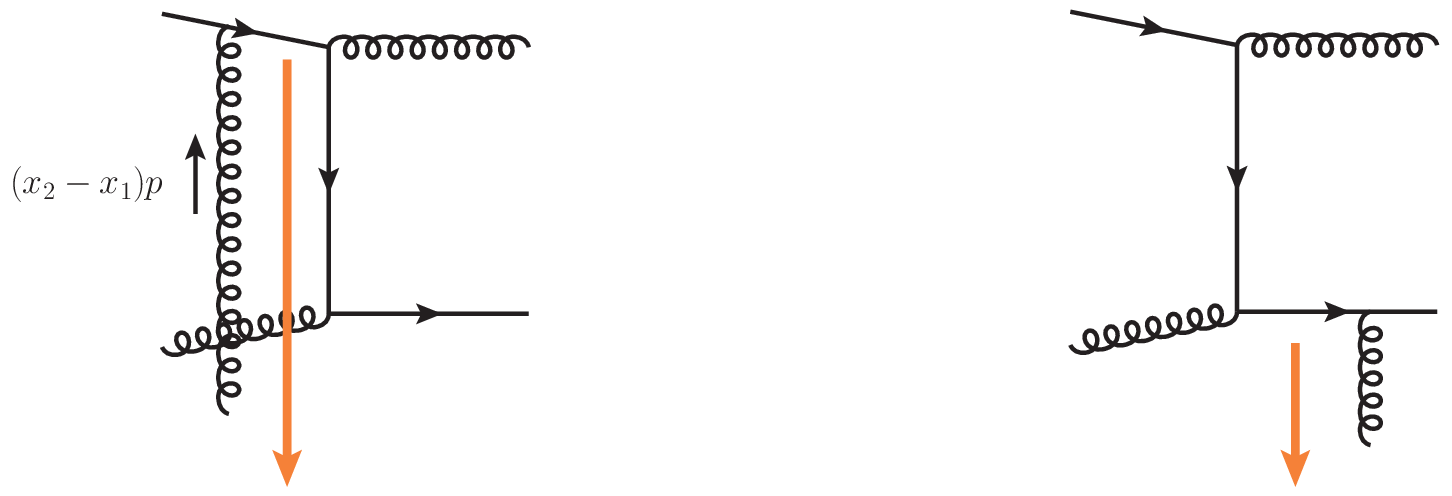}
\end{center}
\end{figure}
\vspace{-15mm}
\beq
&&{1\over (x'p'+(x_2-x_1)p)^2+i\epsilon}\hspace{25mm}
{1\over ({P_h\over z}-(x_2-x_1)p)^2+i\epsilon}
\nonumber\\
&=&{1\over 2x'p\cdot p'}{1\over x_2-x_1+i\epsilon}\hspace{25mm}
=-{1\over 2p\cdot {P_h\over z}}{1\over x_2-x_1-i\epsilon}
\nonumber
\eeq
Although the WTIs (\ref{WTI1})-(\ref{WTI4}) are common in any collisions,
the pole structures in (\ref{SIDIS_Ward1})-(\ref{SIDIS_Ward3}), and then in  (\ref{formula_SIDIS}),
are not uniquely determined from those WTIs in $pp$ collisions.
We provide further discussion on the WTIs.
To illustrate the essence of our discussion, we first consider the following set of the diagrams.
\begin{figure}[H]
\begin{center}
  \includegraphics[height=4cm,width=16cm]{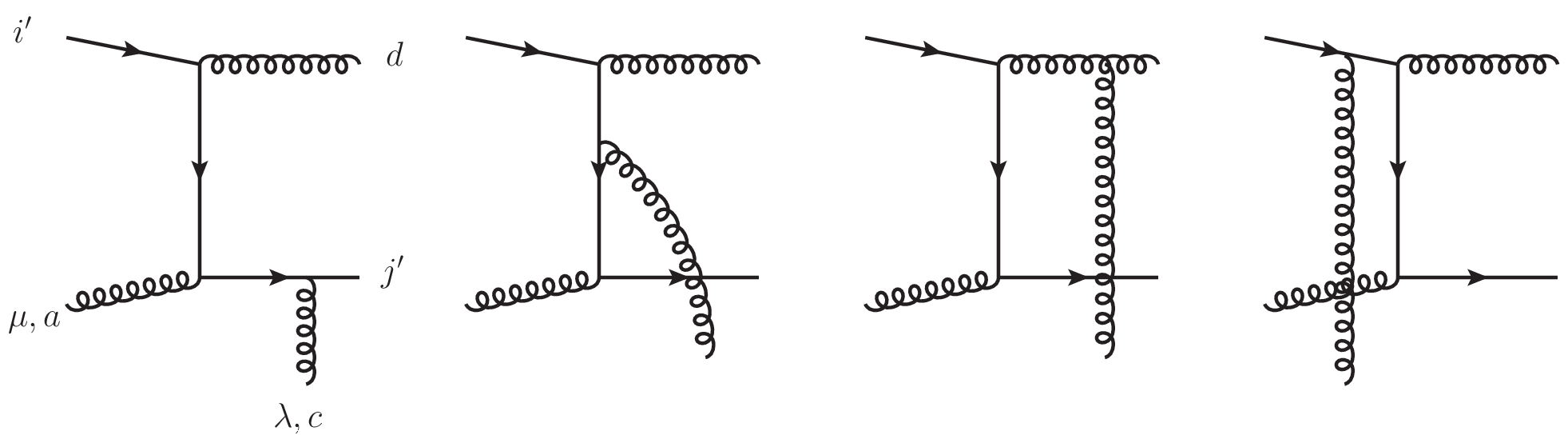}
\end{center}
\end{figure}
\vspace{-12mm}
\beq
\hspace{15mm}
(T^cT^aT^d)_{j'i'}\hspace{21mm}
(T^aT^cT^d)_{j'i'}\hspace{19mm}
(-if^{ced})(T^aT^e)_{j'i'}\hspace{19mm}
(T^aT^dT^c)_{j'i'}
\nonumber
\eeq
$T^a$ is a generator of the color SU($N_c$) group. These color structures are important for our discussion.
We categorize these diagrams into the ISI subset and the FSI subset.
If the gluon line with momentum $k_2-k_1$ is connected to one of the initial(final) external lines,
the diagram belongs to the ISI(FSI) subset. In the present case,
${\cal M}_{4L\mu\lambda}$ belongs to the ISI subset and ${\cal M}_{1L\mu\lambda}$ and 
${\cal M}_{3L\mu\lambda}$ belong to the FSI subset.
If the gluon line is connected to one of internal propagators, the diagram could belong to
both subsets depending on the structure of its color factor. 
The color factor of ${\cal M}_{2L\mu\lambda}$ can be decomposed as
$(T^aT^cT^d)_{j'i'}=(T^aT^dT^c)_{j'i'}+(-if^{ced})(T^aT^e)_{j'i'}$, which is the combination of the 
color factors of the ISI diagram ${\cal M}_{4L\mu\lambda}$ and 
the FSI diagram ${\cal M}_{3L\mu\lambda}$.
We therefore construct two subsets as follows:
\beq
&&{\rm ISI\ subset:}\ (T^aT^dT^c)_{j'i'}({\cal M}_{2L\mu\lambda}+{\cal M}_{4L\mu\lambda})
\nonumber\\
&&{\rm FSI\ subset:}\ (T^cT^aT^d)_{j'i'}{\cal M}_{1L\mu\lambda}
+(-if^{ced})(T^aT^e)_{j'i'}({\cal M}_{2L\mu\lambda}+{\cal M}_{3L\mu\lambda})
\nonumber
\eeq
Performing the same procedure for the remaining diagrams in
Fig. \ref{SL_amp}, we obtain the full result of the ISI subset ${\cal M}_{L\mu\lambda}^{[+]ac}$
and the FSI subset ${\cal M}_{L\mu\lambda}^{[-]ac}$,
\beq
{\cal M}_{L\mu\lambda}^{[+]ac}(k_1,k_2-k_1)&=&
(T^aT^dT^c)_{j'i'}\Bigl({\cal M}_{2L\mu\lambda}+{\cal M}_{4L\mu\lambda}\Bigr)
+(T^dT^aT^c)_{j'i'}{\cal M}_{8L\mu\lambda}
\nonumber\\
&&+(-if^{ade})(T^eT^c)_{j'i'}\Bigl(-{\cal M}_{10L\mu\lambda}
+{\cal M}_{12L\mu\lambda}+{\cal M}_{13L\mu\lambda}^2\Bigr),
\\
{\cal M}_{L\mu\lambda}^{[-]ac}(k_1,k_2-k_1)&=&
(T^cT^aT^d)_{j'i'}{\cal M}_{1L\mu\lambda}
+(-if^{ced})(T^aT^e)_{j'i'}\Bigl({\cal M}_{2L\mu\lambda}+{\cal M}_{3L\mu\lambda}\Bigr)
\nonumber\\
&&+(T^cT^dT^a)_{j'i'}\Bigl({\cal M}_{5L\mu\lambda}+{\cal M}_{6L\mu\lambda}\Bigr)
+(-if^{ced})(T^eT^a)_{j'i'}\Bigl(-{\cal M}_{6L\mu\lambda}+{\cal M}_{7L\mu\lambda}\Bigr)
\nonumber\\
&&+(-if^{ade})(T^cT^e)_{j'i'}\Bigl({\cal M}_{9L\mu\lambda}
+{\cal M}_{10L\mu\lambda}-{\cal M}_{13L\mu\lambda}^2\Bigr)
\nonumber\\
&&+(-if^{aeg})(-if^{ced})(T^g)_{j'i'}\Bigl({\cal M}_{11L\mu\lambda}
+{\cal M}_{13L\mu\lambda}^1\Bigr),
\eeq
where we decomposed ${\cal M}_{13L\mu\lambda}^{ac}$ including a 4-gluon vertex as
\beq
{\cal M}_{13L\mu\lambda}^{ac}=(-if^{aeg})(-if^{ced})(T^g)_{j'i'}{\cal M}_{13L\mu\lambda}^1
+(-if^{ade})(-if^{ceg})(T^g)_{j'i'}{\cal M}_{13L\mu\lambda}^2.
\eeq
We can derive the WTIs for each subset independently,
\beq
(k_2-k_1)^{\lambda}{\cal M}_{L\mu\lambda}^{[\pm]ac}(k_1,k_2-k_1)
=\sum_{k=1}^3(C^{[\pm]}_{Lk\,ac})_{j'i'}{\cal M}_{k\,\mu}(k_2).
\label{WTI_pp}
\eeq
The coefficients are given by
\beq
&&C^{[+]}_{L1\,ac}=-T^aT^dT^c,\hspace{5mm}
C^{[+]}_{L2\,ac}=-T^dT^aT^c,\hspace{5mm}
C^{[+]}_{L3\,ac}=if^{ade}T^eT^c,
\eeq
\beq
&&C^{[-]}_{L1\,ac}=T^cT^aT^d
+if^{ced}T^aT^e,\hspace{5mm}
C^{[-]}_{L2\,ac}=T^dT^cT^a,\hspace{5mm}
C^{[-]}_{L3\,ac}=(-if^{ade})T^cT^e-(-if^{aeg})(-if^{ced})T^g.
\eeq
These WTIs can be combined as
\beq
(k_2-k_1)^{\lambda}\Bigl({\cal M}_{L\mu\lambda}^{[+]ac}(k_1,k_2-k_1)
+{\cal M}_{L\mu\lambda}^{[-]ac}(k_1,k_2-k_1)\Bigr)
=(-if^{ace})\sum_{k=1}^3(C^e_k)_{j'i'}{\cal M}_{k\,\mu}(k_2).
\eeq
Since ${\cal M}_{L\mu\lambda}^{ac}$ in (\ref{SL}) is given by
${\cal M}_{L\mu\lambda}^{ac}={\cal M}_{L\mu\lambda}^{[+]ac}
+{\cal M}_{L\mu\lambda}^{[-]ac}$, the above WTI is rewritten as a relation
at the level of 
the hard part $S_{L\mu\nu\lambda}^{abc}$,
\beq
(k_2-k_1)^{\lambda}S_{L\mu\nu\lambda}^{abc}(k_1,k_2)
=if^{abc}S_{\mu\nu}(k_2),
\eeq
which is consistent with the general WTI (\ref{WTI2}).
Our result shows that general WTI (\ref{WTI2}) 
can be decomposed into minimal WTIs (\ref{WTI_pp}) in $pp$ collisions.
Further decomposition
of ${\cal M}_{L\mu\lambda}^{[+]ac}(k_1,k_2-k_1)$ and ${\cal M}_{L\mu\lambda}^{[-]ac}(k_1,k_2-k_1)$
is needed with respect to another external gluon line with momentum $k_1$ as
\beq
{\cal M}_{L\mu\lambda}^{[\pm]ac}(k_1,k_2-k_1)
={\cal M}_{L\mu\lambda}^{(+)[\pm]ac}(k_1,k_2-k_1)
+{\cal M}_{L\mu\lambda}^{(-)[\pm]ac}(k_1,k_2-k_1).
\eeq
Performing the above procedure with respect to the gluon line with momentum $k_1$,
we derive the following subsets.
\beq
{\cal M}_{L\mu\lambda}^{(+)[+]ac}&=&0,
\eeq
\beq
{\cal M}_{L\mu\lambda}^{(-)[+]ac}&=&
(T^aT^dT^c)_{j'i'}{\cal M}_{2L\mu\lambda}
+(T^aT^dT^c)_{j'i'}\Bigl({\cal M}_{4L\mu\lambda}+{\cal M}_{8L\mu\lambda}\Bigr)
+(-if^{ade})(T^eT^c)_{j'i'}\Bigl({\cal M}_{8L\mu\lambda}+{\cal M}_{12L\mu\lambda}\Bigr)
\nonumber\\
&&+(-if^{ade})(T^eT^c)_{j'i'}\Bigl(-{\cal M}_{10L\mu\lambda}
+{\cal M}_{13L\mu\lambda}^2\Bigr),
\eeq
\beq
{\cal M}_{L\mu\lambda}^{(+)[-]ac}&=&
(T^cT^dT^a)_{j'i'}\Bigl({\cal M}_{1L\mu\lambda}+{\cal M}_{5L\mu\lambda}\Bigr)
+(T^dT^cT^a)_{j'i'}{\cal M}_{6L\mu\lambda}
\nonumber\\
&&+(-if^{ced})(T^eT^a)_{j'i'}\Bigl({\cal M}_{7L\mu\lambda}+{\cal M}_{11L\mu\lambda}
+{\cal M}_{13L\mu\lambda}^1\Bigr),
\eeq
\beq
{\cal M}_{L\mu\lambda}^{(-)[-]ac}&=&
(-if^{ade})(T^cT^e)_{j'i'}\Bigl(-{\cal M}_{1L\mu\lambda}+{\cal M}_{9L\mu\lambda}\Bigr)
+(-if^{ced})(T^aT^e)_{j'i'}{\cal M}_{2L\mu\lambda}
\nonumber\\
&&+(-if^{ced})(T^aT^e)_{j'i'}\Bigl({\cal M}_{3L\mu\lambda}
-{\cal M}_{11L\mu\lambda}-{\cal M}_{13L\mu\lambda}^1\Bigr)
+(-if^{ade})(T^cT^e)_{j'i'}\Bigl({\cal M}_{10L\mu\lambda}-{\cal M}_{13L\mu\lambda}^2\Bigr).
\hspace{5mm}
\eeq
We can show that these subsets independently satisfy the following WTIs with respect to both
$k_2-k_1$ and $k_1$ gluon lines,
\beq
(k_2-k_1)^{\lambda}{\cal M}_{L\mu\lambda}^{(\pm)[\pm]ac}(k_1,k_2-k_1)
&=&\sum_{k=1}^{3} (C^{(\pm)[\pm]}_{Lk\,ac})_{j`i'}{\cal M}_{k\,\mu}(k_2),
\label{WTI_pp1}
\\
k_1^{\mu}{\cal M}_{L\mu\lambda}^{(\pm)[\pm]ac}(k_1,k_2-k_1)&=&
-\sum_{k=1}^{3} (C^{(\pm)[\pm]}_{Lk\,ac})_{j'i'}{\cal M}_{k\,\lambda}(k_2).
\label{WTI_pp2}
\eeq
The coefficients are given by
\beq
C_{L1\,ac}^{(-)[+]}=-T^aT^dT^c,\hspace{5mm}
C_{L2\,ac}^{(-)[+]}=-T^dT^aT^c,\hspace{5mm}
C_{L3\,ac}^{(-)[+]}=-(-if^{ade})T^eT^c,
\eeq
\beq
C_{Lk\,ac}^{(-)[+]}=-C_{Lk\,ca}^{(+)[-]},
\label{WTI_color2}
\eeq
\beq
C^{(-)[-]}_{L1\,ac}&=&-(-if^{ade})T^cT^e-(-if^{ced})T^aT^e,\hspace{5mm}
C^{(-)[-]}_{L2\,ac}=0,\hspace{5mm}
C^{(-)[-]}_{L3\,ac}=(-if^{ade})T^cT^e+(-if^{ced})T^aT^e.
\eeq
These relations give the WTIs at the level of the hard part $S_{L\mu\nu\lambda}^{abc}$,
\beq
(k_2-k_1)^{\lambda}S_{L\mu\nu\lambda}^{(\pm)[\pm]abc}(k_1,k_2)
&=&\sum_{k,l=1}^{3} C^{(\pm)[\pm]}_{Lkl\,abc}S_{kl\,\mu\nu}(k_2),
\\
k_1^{\mu}S_{L\mu\nu\lambda}^{(\pm)[\pm]abc}(k_1,k_2)
&=&-\sum_{k,l=1}^{3} C^{(\pm)[\pm]}_{Lkl\,abc}S_{kl\,\lambda\nu}(k_2).
\eeq
where we defined
\beq
S_{L\mu\nu\lambda}^{(\pm)[\pm]abc}(k_1,k_2)
&=&{\rm Tr}[{\cal M}^{(\pm)[\pm]ac}_{L\mu\lambda}(k_1,k_2-k_1)
(\sum_k C^b_k{\cal M}_{k\,\nu}(k_2))^\dagger]\delta\Bigl((k_2+x'p'-{P_h\over z})^2\Bigr),
\\
S_{kl\,\mu\nu}(k_2)&=&{\rm Tr}[{\cal M}_{k\,\mu}(k_2)
{\cal M}_{l\,\nu}^{\dagger}(k_2)]\delta\Bigl((k_2+x'p'-{P_h\over z})^2\Bigr),\hspace{5mm}
C^{(\pm)[\pm]}_{Lkl\,abc}={\rm Tr}[C^{(\pm)[\pm]}_{Lk\,ac}C^{b\dagger}_l].
\eeq
The color factors $C^{(\pm)[\pm]}_{Lkl\,abc}$ are decomposed 
in terms of SU($N_c$) structure constants as
\beq
C^{(\pm)[\pm]}_{Lkl\, abc}&=&C^{f(\pm)[\pm]}_{Lkl}if^{abc}+C^{d(\pm)[\pm]}_{Lkl}d^{abc}.
\eeq
The condition (\ref{WTI_color2}) gives
\beq
C^{f(-)[+]}_{Lkl}=C^{f(+)[-]}_{Lkl},\hspace{5mm}C^{d(-)[+]}_{Lkl}=-C^{d(+)[-]}_{Lkl}.
\label{WTI_color3}
\eeq
One can calculate all coefficients as follows:
\beq
C^{f(-)[+]}_{Lkl}=\left(
  \begin{array}{ccc}
      \hspace{2mm}{C_F\over 4N_c}-{1\over 8} \hspace{2mm} & 
      \hspace{2mm}{C_F\over 4N_c}-{1\over 8}\hspace{2mm}
       & \hspace{2mm}0\hspace{2mm}  \\ \\
      {C_F\over 4N_c}-{1\over 8} & {C_F\over 4N_c} & {1\over 8} \\ \\
      0 & {1\over 8} & {1\over 8}
    \end{array}
  \right),\hspace{5mm}
  C^{f(-)[-]}_{Lkl}=\left(
  \begin{array}{ccc}
      \hspace{2mm}{1\over 4} \hspace{2mm} & 
      \hspace{2mm}0\hspace{2mm}
       & \hspace{2mm}-{1\over 4}\hspace{2mm}  \\ \\
      0 & 0 & 0 \\ \\
      -{1\over 4} & 0 & {1\over 4}
    \end{array}
  \right),
\eeq
\beq
C^{d(-)[+]}_{Lkl}=\left(
  \begin{array}{ccc}
      \hspace{2mm}-{C_F\over 4N_c}+{1\over 8} \hspace{2mm} & 
      \hspace{2mm}-{C_F\over 4N_c}+{1\over 8}\hspace{2mm}
       & \hspace{2mm}0\hspace{2mm}  \\ \\
      -{C_F\over 4N_c}+{1\over 8} & -{C_F\over 4N_c} & -{1\over 8} \\ \\
      0 & -{1\over 8} & -{1\over 8}
    \end{array}
  \right),\hspace{5mm}
  C^{d(-)[-]}_{Lkl}=0.
\eeq
We find relations
\beq
C^{f(-)[+]}_{Lkl}+C^{f(+)[-]}_{Lkl}+C^{f(-)[-]}_{Lkl}&=&C_{kl},
\label{color_relation1}\\
C^{f,d(\pm)[\pm]}_{L1l}-C^{f,d(\pm)[\pm]}_{L2l}+C^{f,d(\pm)[\pm]}_{L3l}&=&0.
\label{color_relation2}
\eeq
There is an additional WTI
\beq
k_2^{\nu}S_{L\mu\nu\lambda}^{(\pm)[\pm]abc}(k_1,k_2)&=&0,
\label{WTI_pp3}
\eeq
which follows from the tree-level WTI $k^{\nu}(\sum_k C^b_k{\cal M}_{k\,\nu}(k))=0$.
We can directly check another WTI
\beq
k_2^{\mu}\Bigl(\sum_{k,l=1}^{3} C^{(\pm)[\pm]}_{Lkl\,abc}S_{kl\,\mu\nu}(k_2)\Bigr)=0,
\label{WTI_pp4}
\eeq
using the relation (\ref{color_relation2}).
The point of our discussion is that, when we derive the following identities from 
the WTIs (\ref{WTI_pp1}) and (\ref{WTI_pp2}), 
the sign of $i\epsilon$ is uniquely determined within each subset because 
it contains only diagrams that give rise to either $+i\epsilon$ or $-i\epsilon$,
\beq
p^{\lambda}S_{L\mu\nu\lambda}^{(s)[\pm]abc}(k_1,k_2)
&=&{1\over (k_2-k_1)\cdot n\pm i\epsilon}\Bigl(
-\omega^{\lambda}_{\ \gamma}(k_2-k_1)^{\gamma}S_{L\mu\nu\lambda}^{(s)[\pm]abc}(k_1,k_2)
+\sum_{k,l=1}^{3} C^{(s)[\pm]}_{Lkl\,abc}S_{kl\,\mu\nu}(k_2)\Bigr),
\label{pp_Ward1}
\\
p^{\mu}S_{L\mu\nu\lambda}^{(\pm)[s]abc}(k_1,k_2)
&=&{1\over k_1\cdot n\pm i\epsilon}\Bigl(
-\omega^{\mu}_{\ \alpha}k_1^{\alpha}S_{L\mu\nu\lambda}^{(\pm)[s]abc}(k_1,k_2)
-\sum_{k,l=1}^{3} C^{(\pm)[s]}_{Lkl\,abc}S_{kl\,\lambda\nu}(k_2)\Bigr),
\label{pp_Ward2}
\eeq
where the index $s$ takes either $+$ or $-$. 
They are our desired results corresponding to (\ref{SIDIS_Ward1}) and (\ref{SIDIS_Ward2})
in $pp$ collisions. The ambiguity in the sign of $i\epsilon$ is removed by using the minimal 
WTIs for the subsets. We will use the following identities derived from
(\ref{WTI_pp3}) and (\ref{WTI_pp4}) in the next subsection.
\beq
p^{\nu}S_{L\mu\nu\lambda}^{(\mp)[\pm]abc}(k_1,k_2)
&=&-{1\over k_2\cdot n}\omega^{\nu}_{\ \beta}k_2^{\beta}
S_{L\mu\nu\lambda}^{(\mp)[\pm]abc}(k_1,k_2),
\label{pp_Ward3}
\\
p^{\mu}\Bigl(\sum_{k,l=1}^{3} C^{(\pm)[\pm]}_{Lkl\,abc}S_{kl\,\mu\nu}(k_2)\Bigr)
&=&-{1\over k_2\cdot n}\omega^{\mu}_{\ \alpha}k_2^{\alpha}
\Bigl(\sum_{k,l=1}^{3} C^{(\pm)[\pm]}_{Lkl\,abc}S_{kl\,\mu\nu}(k_2)\Bigr).
\label{pp_Ward4}
\eeq
${\cal M}^{bc}_{R\nu\lambda}(k_2,k_2-k_1)$ in (\ref{SR}) can be decomposed
in the same way,
\beq
{\cal M}^{bc}_{R\nu\lambda}=
{\cal M}^{(+)[+]bc}_{R\nu\lambda}
+{\cal M}^{(-)[+]bc}_{R\nu\lambda}
+{\cal M}^{(+)[-]bc}_{R\nu\lambda}
+{\cal M}^{(-)[-]bc}_{R\nu\lambda},
\eeq
where the superscript $(\pm)$ represents the sign of $i\epsilon$ in $1/(k_2\cdot n\pm i\epsilon)$ pole.
${\cal M}^{bc}_{R\nu\lambda}(k_2,k_2-k_1)$ is simply given by
the complex conjugate of ${\cal M}^{ac}_{L\mu\lambda}(k_1,k_2-k_1)$ with
the replacements $\mu\to\nu$, $a\to b$ and $k_1\leftrightarrow k_2$,
\beq
{\cal M}_{R\nu\lambda}^{(\pm)[\pm]bc}(k_2,k_2-k_1)
=({\cal M}_{L\nu\lambda}^{(\mp)[\pm]bc}(k_2,-k_2+k_1))^*.
\eeq
Note that the pole structure with respect to $k_1$ is changed 
upon taking the complex conjugate as
\beq
{1\over k_1\cdot n\pm i\epsilon}\hspace{5mm}\to\hspace{5mm}{1\over k_2\cdot n\mp i\epsilon}.
\eeq
The WTIs are given by
\beq
(k_2-k_1)^{\lambda}{\cal M}_{R\nu\lambda}^{(\pm)[\pm]bc}
=-\sum_{l=1}^3(C^{(\mp)[\pm]}_{Ll\,bc})_{j'i'}^*{\cal M}^*_{l\,\nu}(k_1),\hspace{5mm}
k_2^{\nu}{\cal M}_{R\nu\lambda}^{(\pm)[\pm]bc}
=-\sum_{l=1}^3(C^{(\mp)[\pm]}_{Ll\,bc})_{j'i'}^*{\cal M}^*_{l\,\lambda}(k_1).
\eeq
Then the WTIs at the level of the hard part $S_{R\mu\nu\lambda}^{abc}$ are given by
\beq
(k_2-k_1)^{\lambda}S_{R\mu\nu\lambda}^{(\pm)[\pm]abc}
=-\sum_{k,l=1}^3(C^{(\mp)[\pm]}_{Lkl\,bac})^*S_{kl\,\mu\nu}(k_1),\hspace{5mm}
k_2^{\nu}S_{R\mu\nu\lambda}^{(\pm)[\pm]abc}
=-\sum_{k,l=1}^3(C^{(\mp)[\pm]}_{Lkl\,bac})^*S_{kl\,\mu\lambda}(k_1),
\eeq
where we defined
\beq
S_{R\mu\nu\lambda}^{(\pm)[\pm]abc}
&=&{\rm Tr}[(\sum_k C^a_k{\cal M}_{k\,\mu}(k_1))
{\cal M}^{(\pm)[\pm]bc}_{R\nu\lambda}(k_2,k_2-k_1)]\delta\Bigl((k_1+x'p'-{P_h\over z})^2\Bigr).
\eeq
We find
\beq
(C^{(\pm)[\pm]}_{Lkl\,bac})^*=C^{f(\pm)[\pm]}_{Lkl}(if^{bac})^{*}+C^{d(\pm)[\pm]}_{Lkl}d^{bac}
=C^{(\pm)[\pm]}_{Lkl\,abc}.
\eeq
We can derive the counterparts of (\ref{SIDIS_Ward1}) and (\ref{SIDIS_Ward3}) in $pp$ collisions,
\beq
p^{\lambda}S_{R\mu\nu\lambda}^{(\mp)[\pm]abc}
&=&{1\over (k_2-k_1)\cdot n\pm i\epsilon}\Bigl(
-\omega^{\lambda}_{\ \gamma}(k_2-k_1)^{\gamma}S_{R\mu\nu\lambda}^{(\mp)[\pm]abc}
-\sum_{k,l=1}^3C^{(\pm)[\pm]}_{Lkl\,abc}S_{kl\,\mu\nu}(k_1)\Bigr),
\label{pp_Ward5}
\nonumber\\
p^{\nu}S_{R\mu\nu\lambda}^{(\mp)[\pm]abc}
&=&{1\over k_2\cdot n\mp i\epsilon}\Bigl(
-\omega^{\nu}_{\ \beta}k_1^{\beta}S_{R\mu\nu\lambda}^{(\mp)[\pm]abc}
-\sum_{k,l=1}^3C^{(\pm)[\pm]}_{Lkl\,abc}S_{kl\,\mu\lambda}(k_1)\Bigr).
\label{pp_Ward6}
\eeq
These identities are required to derive the twist-3 gluon distribution contribution to the 
polarized cross section, which is discussed in the next subsection.

\subsection{Derivation of the polarized cross section formula}

We derive the result corresponding to (\ref{formula_SIDIS}) for $pp$ collisions
based on the WTIs derived in the previous subsection.
We first consider contributions from $w^{\rm Fig.1left}$ in (\ref{hadronic})
up to twist-3. We use the trick introduced in \cite{Beppu:2010qn},
\beq
\la pS|A^{\nu}(0)A^{\mu}(\xi)|pS\ra S_{\mu\nu}(k)
&=&\la pS|A^{\beta}(0)A^{\alpha}(\xi)|pS\ra(p^{\nu}n_{\beta}+\omega^{\nu}_{\ \beta})
(p^{\mu}n_{\alpha}+\omega^{\mu}_{\ \alpha})S_{\mu\nu}(k)
\nonumber\\
&=&{1\over (k\cdot n)^2}\omega^{\mu}_{\ \alpha}\omega^{\nu}_{\ \beta}
\la pS|\Bigl(k^{\beta}A^n(0)-k\cdot n A^{\beta}(0)\Bigr)
\Bigl(k^{\alpha}A^n(\xi)-k\cdot n A^{\alpha}(\xi)\Bigr)|pS\ra S_{\mu\nu}(k),\hspace{5mm}
\label{trick}
\eeq
where we used the shorthand notation $A^n=A^{\alpha}n_{\alpha}$ and the identities (\ref{Ward1}).
Integrating by parts with respect to $k$-integral gives
\beq
\Bigl(k^{\alpha}A^n(\xi)-k\cdot n A^{\alpha}(\xi)\Bigr)\hspace{5mm}&\to&\hspace{5mm}
i\Bigl(\partial^{\alpha}A^n(\xi)-\partial^n A^{\alpha}(\xi)\Bigr)=iF^{(0)\alpha n}(\xi),
\\
\Bigl(k^{\beta}A^n(0)-k\cdot n A^{\beta}(0)\Bigr)\hspace{5mm}&\to&\hspace{5mm}
-i\Bigl(\partial^{\beta}A^n(0)-\partial^n A^{\beta}(0)\Bigr)=-iF^{(0)\beta n}(0).
\eeq
Note that the translation invariance shows $\la pS|A^{\beta}(0)A^{\alpha}(\xi)|pS\ra
=\la pS|A^{\beta}(-\xi)A^{\alpha}(0)|pS\ra$.
Then we obtain
\beq
w^{\rm Fig.1left}(p,x'p',{P_h\over z})=
\omega^{\mu}_{\ \alpha}\omega^{\nu}_{\ \beta}\int d^4\xi\int{d^4k\over (2\pi)^4}\,e^{ik\cdot \xi}{1\over (k\cdot n)^2}\la pS|
F^{(0)\beta n}(0)F^{(0)\alpha n}(\xi)|pS\ra S_{\mu\nu}(k).
\label{fig1left}
\eeq
We perform the collinear expansion up to the first derivative term,
\beq
S_{\mu\nu}(k)\simeq S_{\mu\nu}((k\cdot n)p)
+{\partial\over \partial k^{\lambda}}S_{\mu\nu}(k)\Bigr|_{k=(k\cdot n)p}
\omega^{\lambda}_{\ \gamma}k^{\gamma},
\eeq
which shows that the contributions from $w^{\rm Fig.1left}(p,x'p',{P_h\over z})$ up to twist-3
take the form of
\beq
w^{\rm Fig.1left}(p,x'p',{P_h\over z})\Bigr|_{\rm up\ to\ twist-3}
=\omega^{\mu}_{\ \alpha}\omega^{\nu}_{\ \beta}\int {dx\over x^2}\,I^{\alpha\beta}S_{\mu\nu}(xp)
+\omega^{\mu}_{\ \alpha}\omega^{\nu}_{\ \beta}\omega^{\lambda}_{\ \gamma}
\int {dx\over x^2}\,K^{\alpha\beta\gamma}{\partial\over \partial k^{\lambda}}S_{\mu\nu}(k)\Bigr|_{k=xp}.
\label{result_fig1left}
\eeq
The intrinsic part $I^{\alpha\beta}$ and the kinematical part $K^{\alpha\beta\gamma}$ 
are derived from (\ref{fig1left}) by
\beq
I^{\alpha\beta}&=&\int{d\lambda\over 2\pi}\,e^{i\lambda x}\la pS|
F^{(0)\beta n}(0)F^{(0)\alpha n}(\lambda n)|pS\ra,
\label{intrinsic1}
\\
K^{\alpha\beta\gamma}&=&i\int{d\lambda\over 2\pi}\,e^{i\lambda x}\la pS|
F^{(0)\beta n}(0)\partial^{\gamma}F^{(0)\alpha n}(\lambda n)|pS\ra.
\label{kinematical1}
\eeq
We next consider the contributions from $w^{\rm Fig.1right}$ in  (\ref{hadronic}).
The hard part is decomposed into $S_{L\mu\nu\lambda}^{abc}$ and 
$S_{R\mu\nu\lambda}^{abc}$, and they are further decomposed as
\beq
S_{L(R)\mu\nu\lambda}^{abc}=S_{L(R)\mu\nu\lambda}^{(+)[+]abc}
+S_{L(R)\mu\nu\lambda}^{(-)[+]abc}+S_{L(R)\mu\nu\lambda}^{(+)[-]abc}
+S_{L(R)\mu\nu\lambda}^{(-)[-]abc}.
\eeq
We first focus on $S_{L\mu\nu\lambda}^{abc}$ part.
We use the same trick as (\ref{trick}) in the following steps.

\

\noindent
\underline{Step1}
\beq
A^{\nu}_b(0)S^{(\mp)[\pm]abc}_{L\mu\nu\lambda}(k_1,k_2)
=-{1\over k_2\cdot n}\omega^{\nu}_{\ \beta}\Bigl(k_2^\beta A^n_b(0)
-k_2\cdot n A^\beta_b(0)\Bigr)S^{(\mp)[\pm]abc}_{L\mu\nu\lambda}(k_1,k_2),
\eeq
\noindent
\underline{Step2}
\beq
A^{\lambda}_c(\eta)S^{(\mp)[\pm]abc}_{L\mu\nu\lambda}(k_1,k_2)
&=&-{1\over (k_2-k_1)\cdot n\pm i\epsilon}\omega^{\lambda}_{\ \gamma}
\Bigl((k_2-k_1)^\gamma A^n_c(\eta)
-(k_2-k_1)\cdot n A^\gamma_c(\eta)\Bigr)S^{(\mp)[\pm]abc}_{L\mu\nu\lambda}(k_1,k_2)
\nonumber\\
&&+{1\over (k_2-k_1)\cdot n\pm i\epsilon}A^n_c(\eta)
\sum_{k,l=1}^{3} C^{(\mp)[\pm]}_{Lkl\,abc}S_{kl\,\mu\nu}(k_2),
\eeq
\noindent
\underline{Step3}
\beq
A^{\mu}_a(\xi)S^{(\mp)[\pm]abc}_{L\mu\nu\lambda}(k_1,k_2)
&=&-{1\over k_1\cdot n\mp i\epsilon}\omega^{\mu}_{\ \alpha}
\Bigl(k_1^\alpha A^n_a(\xi)-k_1\cdot n A^\alpha_a(\xi)\Bigr)S^{(\mp)[\pm]abc}_{L\mu\nu\lambda}(k_1,k_2)
\nonumber\\
&&-{1\over k_1\cdot n\mp i\epsilon}A^n_a(\xi)
\sum_{k,l=1}^{3} C^{(\mp)[\pm]}_{Lkl\,abc}S_{kl\,\lambda\nu}(k_2),
\eeq
\beq
A^n_c(\eta)A^{\mu}_a(\xi)\sum_{k,l=1}^{3} C^{(\mp)[\pm]}_{Lkl\,abc}S_{kl\,\mu\nu}(k_2)
&=&-{1\over k_2\cdot n}\omega^{\mu}_{\ \alpha}\Bigr(
k_2^{\alpha}A^n_c(\eta) A^n_a(\xi)
-k_2\cdot n A^n_c(\eta)A^{\alpha}_a(\xi)\Bigr)
\sum_{k,l=1}^{3} C^{(\mp)[\pm]}_{Lkl\,abc}S_{kl\,\mu\nu}(k_2),\hspace{5mm}
\eeq
where we used (\ref{pp_Ward1})-(\ref{pp_Ward4}). Integrating by parts gives
\beq
\Bigl(k_2^\beta A^n_b(0)-k_2\cdot n A^\beta_b(0)\Bigr)\hspace{5mm}&\to&\hspace{5mm}
-iF_b^{(0)\beta n}(0),
\\
\Bigl((k_2-k_1)^\gamma A^n_c(\eta)-(k_2-k_1)\cdot n A^\gamma_c(\eta)\Bigr)
\hspace{5mm}&\to&\hspace{5mm}iF_c^{(0)\gamma n}(\eta),
\\
\Bigl(k_1^\alpha A^n_a(\xi)-k_1\cdot n A^\alpha_a(\xi)\Bigr)\hspace{5mm}&\to&\hspace{5mm}
iF_a^{(0)\alpha n}(\xi),
\\
\Bigr(k_2^{\alpha}A^n_c(\eta) A^n_a(\xi)-k_2\cdot n A^n_c(\eta)A^{\alpha}_a(\xi)\Bigr)
\hspace{5mm}&\to&\hspace{5mm}
iF_c^{(0)\alpha n}(\eta)A_a^n(\xi)+iA^n_c(\eta)F_a^{(0)\alpha n}(\xi)
\nonumber\\
&&\hspace{5mm}
+(k_2-k_1)\cdot n \Bigl(A_c^{\alpha}(\eta)A_a^{n}(\xi)-A_c^{n}(\eta)A_a^{\alpha}(\xi)\Bigr).
\eeq
$S_{L\mu\nu\lambda}^{(+)[+]abc}$ is absent in the present case.
$S_{L\mu\nu\lambda}^{(-)[-]abc}$ gives,
\beq
&&w^{\rm Fig.1right}(p,x'p',{P_h\over z})\Bigr|^{S^{(-)[-]}_L}
\nonumber\\
&=&{1\over 2}\int d^4\xi\int d^4\eta\int{d^4k_1\over (2\pi)^4}\int{d^4k_2\over (2\pi)^4}
\,e^{ik_1\cdot\xi}e^{i(k_2-k_1)\cdot\eta}{1\over k_2\cdot n}
\nonumber\\
&&\times\Bigl\{-i\omega^{\mu}_{\ \alpha}\omega^{\nu}_{\ \beta}\omega^{\lambda}_{\ \gamma}
{1\over k_1\cdot n-i\epsilon}{1\over k_2\cdot n-k_1\cdot n- i\epsilon}
\la pS|F_b^{(0)\beta n}(0)gF^{(0)\gamma n}_c(\eta)
F^{(0)\alpha n}_a(\xi)|pS\ra S_{L\mu\nu\lambda}^{(-)[-]abc}(k_1,k_2)
\nonumber\\
&&+\omega^{\mu}_{\ \alpha}\omega^{\nu}_{\ \beta}{1\over k_2\cdot n}
\Bigl[-{1\over k_1\cdot n-i\epsilon}
\la pS|F_b^{(0)\beta n}(0)F^{(0)\alpha n}_c(\eta)gA^n_a(\xi)|pS\ra
\nonumber\\
&&+{1\over k_2\cdot n-k_1\cdot n- i\epsilon}
\la pS|F_b^{(0)\beta n}(0)gA^{n}_c(\eta)F^{(0)\alpha n}_a(\xi)|pS\ra
\nonumber\\
&&-i
\la pS|F_b^{(0)\beta n}(0)g\Bigl(A_c^{\alpha}(\eta)A_a^{n}(\xi)-A_a^{\alpha}(\eta)A_c^{n}(\xi)\Bigr)|pS\ra
\Bigr]\sum_{k,l=1}^{3} C^{(-)[-]}_{Lkl\,abc}S_{kl\,\mu\nu}(k_2)
\Bigr\}.
\eeq
We find that the $d^{abc}$ part of the color factor $C^{(-)[-]}_{Lkl\,abc}$ cancels even if it is nonzero,
\beq
&&{1\over 2}\int d^4\xi\int d^4\eta\int{d^4k_1\over (2\pi)^4}\int{d^4k_2\over (2\pi)^4}
\,e^{ik_1\cdot\xi}e^{i(k_2-k_1)\cdot\eta}{1\over (k_2\cdot n)^2}
\nonumber\\
&&\times \omega^{\mu}_{\ \alpha}\omega^{\nu}_{\ \beta}
\Bigl[{1\over k_2\cdot n-k_1\cdot n- i\epsilon}
\la pS|F_b^{(0)\beta n}(0)\Bigl(gA^{n}_c(\eta)F^{(0)\alpha n}_a(\xi)
-gA^n_a(\eta)F^{(0)\alpha n}_c(\xi)\Bigr)|pS\ra
\nonumber\\
&&-i\la pS|F_b^{(0)\beta n}(0)g\Bigl(A_c^{\alpha}(\eta)A_a^{n}(\xi)-A_a^{\alpha}(\eta)A_c^{n}(\xi)\Bigr)
|pS\ra\Bigr]d^{abc}\sum_{k,l=1}^{3} C^{d(-)[-]}_{Lkl}S_{kl\,\mu\nu}(k_2)=0,
\eeq
by appropriately changing the integration variables $k_1\leftrightarrow k_2-k_1$ and $\xi\leftrightarrow\eta$.
Then we obtain
\beq
&&w^{\rm Fig.1right}(p,x'p',{P_h\over z})\Bigr|^{S^{(-)[-]}_L}
\nonumber\\
&=&{1\over 2}\omega^{\mu}_{\ \alpha}\omega^{\nu}_{\ \beta}
\int d^4\xi\int d^4\eta\int{d^4k_1\over (2\pi)^4}\int{d^4k_2\over (2\pi)^4}
\,e^{ik_1\cdot\xi}e^{i(k_2-k_1)\cdot\eta}{1\over k_2\cdot n}
\nonumber\\
&&\times\Bigl\{-i\omega^{\lambda}_{\ \gamma}{1\over k_1\cdot n-i\epsilon}
{1\over k_2\cdot n-k_1\cdot n- i\epsilon}
\la pS|F_b^{(0)\beta n}(0)gF^{(0)\gamma n}_c(\eta)
F^{(0)\alpha n}_a(\xi)|pS\ra S_{L\mu\nu\lambda}^{(-)[-]abc}(k_1,k_2)
\nonumber\\
&&+if^{abc}\Bigl[2{1\over k_2\cdot n}{1\over k_2\cdot n-k_1\cdot n- i\epsilon}
\la pS|F_b^{(0)\beta n}(0)gA^n_c(\eta)F^{(0)\alpha n}_a(\xi)|pS\ra
\nonumber\\
&&-2i{1\over k_2\cdot n}\la pS|F_b^{(0)\beta n}(0)gA_c^{\alpha}(\eta)A_a^{n}(\xi)|pS\ra\Bigr]
\sum_{k,l=1}^{3} C^{f(-)[-]}_{Lkl}S_{kl\,\mu\nu}(k_2)
\Bigr\}.
\label{SL--}
\eeq
We perform the collinear expansion
\beq
S_{kl\,\mu\nu}(k_2)&\simeq&S_{kl\,\mu\nu}((k_2\cdot n)p)
+{\partial\over \partial k_2^{\lambda}}S_{kl\,\mu\nu}(k_2)\Bigr|_{k_2=x_2p},
\\
S_{L\mu\nu\lambda}^{abc}(k_1,k_2)&\simeq&
S_{L\mu\nu\lambda}^{abc}((k_1\cdot n)p,(k_2\cdot n)p).
\eeq
The contributions 
from $w^{\rm Fig.1right}(p,x'p',{P_h\over z})$ up to twist-3
are written in the following form.
\beq
w^{\rm Fig.1right}(p,x'p',{P_h\over z})\Bigr|_{\rm up\ to\ twist-3}
&=&\omega^{\mu}_{\ \alpha}\omega^{\nu}_{\ \beta}\sum_{k,l=1}^3
\int {dx\over x^2}I_{kl}^{\alpha\beta}S_{kl\,\mu\nu}(xp)
\nonumber\\
&&+\omega^{\mu}_{\ \alpha}\omega^{\nu}_{\ \beta}\omega^{\lambda}_{\ \gamma}
\sum_{k,l=1}^3\int {dx\over x^2}K_{kl}^{\alpha\beta\gamma}
{\partial\over \partial k^{\lambda}}S_{kl\,\mu\nu}(k)\Bigr|_{k=xp}
\nonumber\\
&&-{1\over 2}\omega^{\mu}_{\ \alpha}\omega^{\nu}_{\ \beta}\omega^{\lambda}_{\ \gamma}
\int dx_1\int {dx_2\over x_2}\,
D^{\alpha\beta\gamma}{1\over x_1\mp i\epsilon}{1\over x_2-x_1\pm i\epsilon}
S_{L\mu\nu\lambda}^{(\mp)[\pm]abc}(x_1p,x_2p)
\nonumber\\
&&-{1\over 2}\omega^{\mu}_{\ \alpha}\omega^{\nu}_{\ \beta}\omega^{\lambda}_{\ \gamma}
\int {dx_1\over x_1}\int {dx_2}\,
D^{\alpha\beta\gamma}{1\over x_2\mp i\epsilon}{1\over x_2-x_1\pm i\epsilon}
S_{R\mu\nu\lambda}^{(\mp)[\pm]abc}(x_1p,x_2p).\hspace{5mm}
\label{result_SL}
\eeq
The contributions associated with \(S_{L\mu\nu\lambda}^{(-)[-]abc}(k_1,k_2)\) in (\ref{SL--}) are separated into each coefficient as follows:
\beq
I_{kl}^{\alpha\beta}\Bigr|^{S^{(-)[-]}_L}&=&
\int{d\lambda\over 2\pi}
\,e^{i\lambda x}\Bigl[
\la pS|F_b^{(0)\beta n}(0)\Bigl(ig\int^{\infty}_{\lambda}d\mu\,A^n_c(\mu n)if^{abc}
\Bigr)F^{(0)\alpha n}_a(\lambda n)|pS\ra
\nonumber\\
&&+\la pS|F_b^{(0)\beta n}(0)\Bigl(gf^{abc}A_c^{\alpha}(\lambda n)A_a^{n}(\lambda n)\Bigr)|pS\ra
\Bigr]C^{f(-)[-]}_{Lkl},
\label{intrinsic2}
\eeq
\beq
K_{kl}^{\alpha\beta\gamma}\Bigr|^{S^{(-)[-]}_L}&=&
i\int{d\lambda\over 2\pi}\,e^{i\lambda x}\Bigl[
\la pS|F_b^{(0)\beta n}(0)\Bigl(ig\int^{\infty}_{\lambda}d\mu\,A^n_c(\mu n)if^{abc}
\Bigr)\partial^{\gamma}F^{(0)\alpha n}_a(\lambda n)|pS\ra
\nonumber\\
&&+ig\int^{\infty}_{\lambda}d\mu\,\la pS|F_b^{(0)\beta n}(0)F^{(0)\gamma n}_c(\mu n)(if^{abc})
F^{(0)\alpha n}_a(\lambda n)|pS\ra
\nonumber\\
&&+\la pS|F_b^{(0)\beta n}(0)\Bigl(-igA_c^{\gamma}(\lambda n)if^{abc}\Bigr)
F^{(0)\alpha n}_a(\lambda n)|pS\ra
\nonumber\\
&&+\la pS|F_b^{(0)\beta n}(0)\partial^{\gamma}
\Bigl(gf^{abc}A_c^{\alpha}(\lambda n)A_a^{n}(\lambda n)\Bigr)
|pS\ra\Bigr]C^{f(-)[-]}_{Lkl},
\label{kinematical2}
\eeq
\beq
D^{\alpha\beta\gamma}&=&
i\int{d\lambda\over 2\pi}\int{d\mu\over 2\pi}e^{i\lambda x_1}e^{i\mu(x_2-x_1)}\la pS|F_b^{(0)\beta n}(0)gF^{(0)\gamma n}_c(\mu n)
F^{(0)\alpha n}_a(\lambda n)|pS\ra,
\eeq
where we used the formula
\beq
\int dx{1\over x-i\epsilon}e^{i\alpha x}=2\pi i\theta(\alpha).
\eeq
It is convenient to consider the $S_{L\mu\nu\lambda}^{(-)[+]abc}(k_1,k_2)$ and 
$S_{L\mu\nu\lambda}^{(+)[-]abc}(k_1,k_2)$ parts together because 
the color factors in their WTIs are not independent as shown in (\ref{WTI_color3}). 
In this case, $d^{abc}$ part is not canceled,
\beq
&&w^{\rm FIG.1right}(p,x'p',{P_h\over z})\Bigr|^{S_{L}^{(-)[+]}+S_L^{(+)[-]}}
\nonumber\\
&=&{1\over 2}\omega^{\mu}_{\ \alpha}\omega^{\nu}_{\ \beta}
\int d^4\xi\int d^4\eta\int{d^4k_1\over (2\pi)^4}\int{d^4k_2\over (2\pi)^4}
\,e^{ik_1\cdot\xi}e^{i(k_2-k_1)\cdot\eta}{1\over k_2\cdot n}
\nonumber\\
&&\times\Bigl\{
-i\omega^{\lambda}_{\ \gamma}\la pS|F_b^{(0)\beta n}(0)gF^{(0)\gamma n}_c(\eta)
F^{(0)\alpha n}_a(\xi)|pS\ra
\Bigl[
{1\over k_1\cdot n-i\epsilon}{1\over k_2\cdot n-k_1\cdot n+i\epsilon}
S_{L\mu\nu\lambda}^{(-)[+]abc}(k_1,k_2)
\nonumber\\
&&+{1\over k_1\cdot n+i\epsilon}{1\over k_2\cdot n-k_1\cdot n-i\epsilon} 
S_{L\mu\nu\lambda}^{(+)[-]abc}(k_1,k_2)
\Bigr]
\nonumber\\
&&+{1\over k_2\cdot n}if^{abc}
\Bigl[2{1\over k_2\cdot n-k_1\cdot n+i\epsilon}
\la pS|F_b^{(0)\beta n}(0)gA^{n}_c(\eta)F^{(0)\alpha n}_a(\xi)|pS\ra
\nonumber\\
&&+2{1\over k_2\cdot n-k_1\cdot n-i\epsilon}
\la pS|F_b^{(0)\beta n}(0)gA^{n}_c(\eta)F^{(0)\alpha n}_a(\xi)|pS\ra
\Bigr]\sum_{k,l=1}^{3} C^{f(-)[+]}_{Lkl}S_{kl\,\mu\nu}(k_2)
\nonumber\\
&&+{1\over k_2\cdot n}d^{abc}
\Bigl[2{1\over k_2\cdot n-k_1\cdot n+i\epsilon}
\la pS|F_b^{(0)\beta n}(0)gA^{n}_c(\eta)F^{(0)\alpha n}_a(\xi)|pS\ra
\nonumber\\
&&-2{1\over k_2\cdot n-k_1\cdot n-i\epsilon}
\la pS|F_b^{(0)\beta n}(0)gA^{n}_c(\eta)F^{(0)\alpha n}_a(\xi)|pS\ra
\Bigr]\sum_{k,l=1}^{3} C^{d(-)[+]}_{Lkl}S_{kl\,\mu\nu}(k_2)
\nonumber\\
&&+2\la pS|F_b^{(0)\beta n}(0)gf^{abc}A_c^{\alpha}(\eta)A_a^{n}(\xi)|pS\ra
\sum_{k,l=1}^{3}\Bigl(C^{f(-)[+]}_{Lkl}+C^{f(+)[-]}_{Lkl}\Bigr)S_{kl\,\mu\nu}(k_2)
\Bigr\}.
\eeq
After the collinear expansion, we obtain
\beq
I_{kl}^{\alpha\beta}\Bigr|^{S_{L}^{(-)[+]}+S_L^{(+)[-]}}&=&
\int{d\lambda\over 2\pi}
\,e^{i\lambda x}\Bigl[
\la pS|F_b^{(0)\beta n}(0)\Bigl(ig\int^{-\infty}_{\lambda}d\mu\,A^n_c(\mu n)if^{abc}
\Bigr)F^{(0)\alpha n}_a(\lambda n)|pS\ra C^{f(-)[+]}_{Lkl}
\nonumber\\
&&+\la pS|F_b^{(0)\beta n}(0)\Bigl(ig\int^{\infty}_{\lambda}d\mu\,A^n_c(\mu n)if^{abc}
\Bigr)F^{(0)\alpha n}_a(\lambda n)|pS\ra C^{f(+)[-]}_{Lkl}
\nonumber\\
&&-\la pS|F_b^{(0)\beta n}(0)\Bigl(ig\int^{\infty}_{-\infty}d\mu\,A^n_c(\mu n)d^{abc}
\Bigr)F^{(0)\alpha n}_a(\lambda n)|pS\ra C^{d(-)[+]}_{Lkl}
\nonumber\\
&&+\la pS|F_b^{(0)\beta n}(0)gf^{abc}A_c^{\alpha}(\lambda n)A_a^{n}(\lambda n)|pS\ra
\Bigl(C^{f(-)[+]}_{Lkl}+C^{f(+)[-]}_{Lkl}\Bigr)\Bigr],
\label{intrinsic3}
\eeq
\beq
K_{kl}^{\alpha\beta\gamma}\Bigr|^{S_{L}^{(-)[+]}+S_L^{(+)[-]}}&=&
i\int{d\lambda\over 2\pi}
\,e^{i\lambda x}\Bigl[
\la pS|F_b^{(0)\beta n}(0)\Bigl(ig\int^{-\infty}_{\lambda}d\mu\,A^n_c(\mu n)if^{abc}
\Bigr)\partial^{\gamma}F^{(0)\alpha n}_a(\lambda n)|pS\ra C^{f(-)[+]}_{Lkl}
\nonumber\\
&&+ig\int^{-\infty}_{\lambda}d\mu\,\la pS|F_b^{(0)\beta n}(0)F^{(0)\gamma n}_c(\mu n)(if^{abc})
F^{(0)\alpha n}_a(\lambda n)|pS\ra C^{f(-)[+]}_{Lkl}
\nonumber\\
&&+\la pS|F_b^{(0)\beta n}(0)\Bigl(ig\int^{\infty}_{\lambda}d\mu\,A^n_c(\mu n)if^{abc}
\Bigr)\partial^{\gamma}F^{(0)\alpha n}_a(\lambda n)|pS\ra C^{f(+)[-]}_{Lkl}
\nonumber\\
&&+ig\int^{\infty}_{\lambda}d\mu\,\la pS|F_b^{(0)\beta n}(0)F^{(0)\gamma n}_c(\mu n)(if^{abc})
F^{(0)\alpha n}_a(\lambda n)|pS\ra C^{f(+)[-]}_{Lkl}
\nonumber\\
&&-\la pS|F_b^{(0)\beta n}(0)\Bigl(ig\int^{\infty}_{-\infty}d\mu\,A^n_c(\mu n)d^{abc}
\Bigr)\partial^{\gamma}F^{(0)\alpha n}_a(\lambda n)|pS\ra C^{d(-)[+]}_{Lkl}
\nonumber\\
&&-ig\int^{\infty}_{-\infty}d\mu\,\la pS|F_b^{(0)\beta n}(0)F^{(0)\gamma n}_c(\mu n)(d^{abc})
F^{(0)\alpha n}_a(\lambda n)|pS\ra C^{d(-)[+]}_{Lkl}
\nonumber\\
&&+\la pS|F_b^{(0)\beta n}(0)\Bigl(-igA_c^{\gamma}(\lambda n)if^{abc}\Bigr)
F^{(0)\alpha n}_a(\lambda n)|pS\ra\Bigl(C^{f(-)[+]}_{Lkl}+C^{f(+)[-]}_{Lkl}\Bigr)
\nonumber\\
&&+\la pS|F_b^{(0)\beta n}(0)\partial^{\gamma}
\Bigl(gf^{abc}A_c^{\alpha}(\lambda n)A_a^{n}(\lambda n)\Bigr)
|pS\ra\Bigl(C^{f(-)[+]}_{Lkl}+C^{f(+)[-]}_{Lkl}\Bigr)
\Bigr].
\label{kinematical3}
\eeq
The contribution from $S_{R\mu\nu\lambda}^{abc}(k_1,k_2)$ can be calculated in the same way.
We use the following tricks.

\

\noindent
\underline{Step1}
\beq
A_{a}^{\mu}(\xi)S_{R\mu\nu\lambda}^{(\mp)[\pm]abc}=
-{1\over k_1\cdot n}\omega^{\mu}_{\ \alpha}
\Bigl(k_1^{\alpha}A_{a}^{n}(\xi)-k_1\cdot n A_{a}^{\alpha}(\xi)\Bigr)
S_{R\mu\nu\lambda}^{(\mp)[\pm]abc},
\eeq

\noindent
\underline{Step2}
\beq
A_{c}^{\lambda}(\eta)S_{R\mu\nu\lambda}^{(\mp)[\pm]abc}&=&
-{1\over (k_2-k_1)\cdot n\pm i\epsilon}\omega^{\lambda}_{\ \gamma}
\Bigl((k_2-k_1)^{\gamma}A_{c}^{n}(\eta)-(k_2-k_1)\cdot n A_{c}^{\gamma}(\eta)\Bigr)
S_{R\mu\nu\lambda}^{(\mp)[\pm]abc}
\nonumber\\
&&-{1\over (k_2-k_1)\cdot n\pm i\epsilon}A_{c}^{n}(\eta)
\sum_{k,l=1}^3C^{(\pm)[\pm]}_{Lkl\,abc}S_{kl\,\mu\nu}(k_1),
\eeq

\noindent
\underline{Step3}
\beq
A_{b}^{\nu}(0)S_{R\mu\nu\lambda}^{(\mp)[\pm]abc}&=&
-{1\over k_2\cdot n\mp i\epsilon}\omega^{\nu}_{\ \beta}
\Bigl(k_2^{\beta}A_{b}^{n}(0)-k_2\cdot n A_{b}^{\beta}(0)\Bigr)
S_{R\mu\nu\lambda}^{(\mp)[\pm]abc}
\nonumber\\
&&-{1\over k_2\cdot n\mp i\epsilon}A^n_b(0)
\sum_{k,l=1}^3C^{(\pm)[\pm]}_{Lkl\,abc}S_{kl\,\mu\lambda}(k_1),
\eeq
\beq
&&A_{b}^{\nu}(0)A_{c}^{n}(\eta)\sum_{k,l=1}^3C^{(\pm)[\pm]}_{Lkl\,bac}
S_{kl\,\mu\nu}(k_1)
\nonumber\\
&=&-{1\over k_1\cdot n}\omega^{\nu}_{\ \beta}
\Bigl(k_1^{\beta}A_{b}^{n}(0)A_{c}^{n}(\eta)
-k_1\cdot n A_{b}^{\beta}(0)A_{c}^{n}(\eta)\Bigr)
\sum_{k,l=1}^3C^{(\pm)[\pm]}_{Lkl\,abc}S_{kl\,\mu\nu}(k_1).
\eeq
After integrating by parts, we obtain
\beq
\Bigl(k_1^{\alpha}A_{a}^{n}(\xi)-k_1\cdot n A_{a}^{\alpha}(\xi)\Bigr)\hspace{5mm}&\to&\hspace{5mm}
iF^{(0)\alpha n}_a(\xi),
\\
\Bigl((k_2-k_1)^{\gamma}A_{c}^{n}(\eta)-(k_2-k_1)\cdot n A_{c}^{\gamma}(\eta)\Bigr)
\hspace{5mm}&\to&\hspace{5mm}
iF^{(0)\gamma n}_c(\eta),
\\
\Bigl(k_2^{\beta}A_{b}^{n}(0)-k_2\cdot n A_{b}^{\beta}(0)\Bigr)\hspace{5mm}&\to&\hspace{5mm}
-iF^{(0)\beta n}_b(0),
\\
\Bigl(k_1^{\beta}A_{b}^{n}(0)A_{c}^{n}(\eta)-k_1\cdot n A_{b}^{\beta}(0)
A_{c}^{n}(\eta)\Bigr)\hspace{5mm}&\to&\hspace{5mm}
-iF^{(0)\beta n}_b(0)A^n_c(\eta)-iA^n_b(0)F^{(0)\beta n}_c(\eta)
\\
&&+(k_2-k_1)\cdot n\Bigl(A^{\beta}_b(0)A^n_c(\eta)-A^{n}_b(0)A^{\beta}_c(\eta)\Bigr).
\eeq
The full results of $I_{kl}^{\alpha\beta}$  and $K_{kl}^{\alpha\beta\gamma}$ are given by
\beq
I_{kl}^{\alpha\beta}\Bigr|^{S_R}&=&\int{d\lambda\over 2\pi}
\,e^{i\lambda x}\Bigl[
-\la pS|F_b^{(0)\beta n}(0)\Bigl(ig\int^{-\infty}_{0}d\mu\,A^n_c(\mu n)if^{abc}
\Bigr)F^{(0)\alpha n}_a(\lambda n)|pS\ra C^{f(-)[+]}_{Lkl}
\nonumber\\
&&-\la pS|F_b^{(0)\beta n}(0)\Bigl(ig\int^{\infty}_{0}d\mu\,A^n_c(\mu n)if^{abc}
\Bigr)F^{(0)\alpha n}_a(\lambda n)|pS\ra \Bigl(C^{f(-)[-]}_{Lkl}+C^{f(+)[-]}_{Lkl}\Bigr)
\nonumber\\
&&+\la pS|F_b^{(0)\beta n}(0)\Bigl(ig\int^{\infty}_{-\infty}d\mu\,A^n_c(\mu n)d^{abc}
\Bigr)F^{(0)\alpha n}_a(\lambda n)|pS\ra C^{d(-)[+]}_{Lkl}
\nonumber\\
&&+\la pS|\Bigl(gf^{abc}A^{\beta}_b(0)A^n_c(0)\Bigr)F_a^{(0)\alpha n}(\lambda n)|pS\ra
\Bigl(C^{f(-)[-]}_{Lkl}+C^{f(+)[-]}_{Lkl}+C^{f(-)[+]}_{Lkl}\Bigr)\Bigr],
\label{intrinsic4}
\eeq
\beq
K_{kl}^{\alpha\beta\gamma}\Bigr|^{S_R}&=&i\int{d\lambda\over 2\pi}
\,e^{i\lambda x}\Bigl[-\la pS|F_b^{(0)\beta n}(0)\Bigl(ig\int^{-\infty}_{0}d\mu\,A^n_c(\mu n)if^{abc}
\Bigr)\partial^{\gamma}F^{(0)\alpha n}_a(\lambda n)|pS\ra C^{f(-)[+]}_{Lkl}
\nonumber\\
&&-\la pS|F_b^{(0)\beta n}(0)\Bigl(ig\int^{\infty}_{0}d\mu\,A^n_c(\mu n)if^{abc}
\Bigr)\partial^{\gamma}F^{(0)\alpha n}_a(\lambda n)|pS\ra \Bigl(C^{f(-)[-]}_{Lkl}+C^{f(+)[-]}_{Lkl}\Bigr)
\nonumber\\
&&+\la pS|F_b^{(0)\beta n}(0)\Bigl(ig\int^{\infty}_{-\infty}d\mu\,A^n_c(\mu n)d^{abc}
\Bigr)\partial^{\gamma}F^{(0)\alpha n}_a(\lambda n)|pS\ra C^{d(-)[+]}_{Lkl}
\nonumber\\
&&+\la pS|\Bigl(gf^{abc}A^{\beta}_b(0)A^n_c(0)\Bigr)
\partial^{\gamma}F_a^{(0)\alpha n}(\lambda n)|pS\ra
\Bigl(C^{f(-)[-]}_{Lkl}+C^{f(+)[-]}_{Lkl}+C^{f(-)[+]}_{Lkl}\Bigr)
\Bigr].
\label{kinematical4}
\eeq
We combine all contribution up to twist-3.
The intrinsic parts (\ref{intrinsic1}), (\ref{intrinsic2}), (\ref{intrinsic3}) and (\ref{intrinsic4}) 
are combined as
\beq
&&I^{\alpha\beta}S_{\mu\nu}(xp)
+\sum_{k,l=1}^3I_{kl}^{\alpha\beta}S_{kl\,\mu\nu}(xp)
\nonumber\\
&=&\int{d\lambda\over 2\pi}\,e^{i\lambda x}\Bigl[\la pS|
F^{(0)\beta n}(0)F^{(0)\alpha n}(\lambda n)|pS\ra
+\la pS|F_b^{(0)\beta n}(0)\Bigl(ig\int^{0}_{\lambda}d\mu\,A^n_c(\mu n)if^{abc}
\Bigr)F^{(0)\alpha n}_a(\lambda n)|pS\ra
\nonumber\\
&&+\la pS|F_b^{(0)\beta n}(0)\Bigl(gf^{abc}A_c^{\alpha}(\lambda n)A_a^{n}(\lambda n)\Bigl)|pS\ra
+\la pS|\Bigl(gf^{abc}A^{\beta}_b(0)A^n_c(0)\Bigr)F_a^{(0)\alpha n}(\lambda n)|pS\ra
\Bigr]S_{\mu\nu}(xp)
\nonumber\\
&=&\Phi^{\alpha\beta}(x)\Bigr|_{{\rm up\ to\ }g^1-{\rm terms}}
S_{\mu\nu}(xp),
\eeq
where we used the identity (\ref{color_relation1}).
We have successfully constructed the gauge-invariant matrix element.
The kinematical parts (\ref{kinematical1}), (\ref{kinematical2}), (\ref{kinematical3}) and (\ref{kinematical4}) 
are combined as
\beq
&&K^{\alpha\beta\gamma}{\partial\over \partial k^{\lambda}}S_{\mu\nu}(k)\Bigr|_{k=xp}
+\sum_{k,l=1}^3K_{kl}^{\alpha\beta\gamma}
{\partial\over \partial k^{\lambda}}S_{kl\,\mu\nu}(k)\Bigr|_{k=xp}
\nonumber\\
&=&i\sum_{k,l=1}^3\int{d\lambda\over 2\pi}\,e^{i\lambda x}\Bigl[
\la pS|F_b^{(0)\beta n}(0)D_{ba}^{\gamma}(\lambda n)F_a^{(0)\alpha n}(\lambda n)|pS\ra
\Bigl(C^{f(-)[-]}_{Lkl}+C^{f(-)[+]}_{Lkl}+C^{f(+)[-]}_{Lkl}\Bigr)
\nonumber\\
&&+\la pS|F_b^{(0)\beta n}(0)\Bigl(ig\int^{0}_{\lambda}d\mu\,A^n_c(\mu n)if^{abc}
\Bigr)\partial^{\gamma}F^{(0)\alpha n}_a(\lambda n)|pS\ra 
\Bigl(C^{f(-)[-]}_{Lkl}+C^{f(-)[+]}_{Lkl}+C^{f(+)[-]}_{Lkl}\Bigr)
\nonumber\\
&&+ig\int^{-\infty}_{\lambda}d\mu\,\la pS|F_b^{(0)\beta n}(0)F^{(0)\gamma n}_c(\mu n)(if^{abc})
F^{(0)\alpha n}_a(\lambda n)|pS\ra C^{f(-)[+]}_{Lkl}
\nonumber\\
&&+ig\int^{\infty}_{\lambda}d\mu\,\la pS|F_b^{(0)\beta n}(0)F^{(0)\gamma n}_c(\mu n)(if^{abc})
F^{(0)\alpha n}_a(\lambda n)|pS\ra \Bigl(C^{f(-)[-]}_{Lkl}+C^{f(+)[-]}_{Lkl}\Bigr)
\nonumber\\
&&-ig\int^{\infty}_{-\infty}d\mu\,\la pS|F_b^{(0)\beta n}(0)F^{(0)\gamma n}_c(\mu n)(d^{abc})
F^{(0)\alpha n}_a(\lambda n)|pS\ra C^{d(-)[+]}_{Lkl}
\nonumber\\
&&+\la pS|F_b^{(0)\beta n}(0)\partial^{\gamma}
\Bigl(gf^{abc}A_c^{\alpha}(\lambda n)A_a^{n}(\lambda n)\Bigr)
|pS\ra\Bigl(C^{f(-)[-]}_{Lkl}+C^{f(-)[+]}_{Lkl}+C^{f(+)[-]}_{Lkl}\Bigr)
\nonumber\\
&&+\la pS|\Bigl(gf^{abc}A^{\beta}_b(0)A^n_c(0)\Bigr)
\partial^{\gamma}F_a^{(0)\alpha n}(\lambda n)|pS\ra
\Bigl(C^{f(-)[-]}_{Lkl}+C^{f(+)[-]}_{Lkl}+C^{f(-)[+]}_{Lkl}\Bigr)
\Bigr]{\partial\over \partial k^{\lambda}}S_{kl\,\mu\nu}(k)\Bigr|_{k=xp}
\nonumber\\
&=&\sum_{k,l=1}^3\int{d\lambda\over 2\pi}\,e^{i\lambda x}\Bigl\{
C^{f(-)[+]}_{Lkl}\Phi^{[+]\alpha\beta\gamma}_{\partial}(x)+\Bigl(C^{f(-)[-]}_{Lkl}+C^{f(+)[-]}_{Lkl}\Bigr)
\Phi^{[-]\alpha\beta\gamma}_{\partial}(x)
\nonumber\\
&&-2\pi i C^{d(-)[+]}_{Lkl}d^{abc}\Phi^{\alpha\beta\gamma}_{F\,abc}(x,x)\Bigr\}
\Bigr|_{{\rm up\ to\ }g^1-{\rm terms}}
{\partial\over \partial k^{\lambda}}S_{kl\,\mu\nu}(k)\Bigr|_{k=xp}.
\eeq
The dynamical parts have the common matrix element $D^{\alpha\beta\gamma}$,
\beq
D^{\alpha\beta\gamma}_{abc}&=&i
\int{d\lambda\over 2\pi}\int{d\mu\over 2\pi}e^{i\lambda x_1}e^{i\mu(x_2-x_1)}
\la pS|F_b^{(0)\beta n}(0)gF^{(0)\gamma n}_c(\mu n)
F^{(0)\alpha n}_a(\lambda n)|pS\ra
\nonumber\\
&=&\Phi^{\alpha\beta\gamma}_{F\,abc}(x_1,x_2)\Bigr|_{{\rm up\ to\ }g^1-{\rm terms}}.
\eeq
We finally obtain the corresponding result to (\ref{formula_SIDIS}) in $pp$ collisions,
\beq
w(p,x'p',{P_h\over z})&=&
\omega^{\mu}_{\ \alpha}\omega^{\nu}_{\ \beta}\int{dx\over x^2}
\Phi^{\alpha\beta}(x)S_{\mu\nu}(xp)
+\omega^{\mu}_{\ \alpha}\omega^{\nu}_{\ \beta}\omega^{\lambda}_{\ \gamma}
\int{dx\over x^2}\Bigl[C^{f(-)[+]}_{Lkl}\Phi^{[+]\alpha\beta\gamma}_{\partial}(x)
\nonumber\\
&&+\Bigl(C^{f(-)[-]}_{Lkl}+C^{f(+)[-]}_{Lkl}\Bigr)
\Phi^{[-]\alpha\beta\gamma}_{\partial}(x)
-2\pi i C^{d(-)[+]}_{Lkl}d^{abc}\Phi^{\alpha\beta\gamma}_{F\,abc}(x,x)
\Bigr]{\partial\over \partial k^{\lambda}}S_{kl\,\mu\nu}(k)\Bigr|_{k=xp}
\nonumber\\
&&-{1\over 2}\omega^{\mu}_{\ \alpha}\omega^{\nu}_{\ \beta}\omega^{\lambda}_{\ \gamma}
\Bigl(\int dx_1\int {dx_2\over x_2}\,\Phi^{\alpha\beta\gamma}_{F\,abc}(x_1,x_2)
{1\over x_1\mp i\epsilon}{1\over x_2-x_1\pm i\epsilon}S^{(\mp)[\pm]abc}_{L\mu\nu\lambda}(x_1p,x_2p)
\nonumber\\
&&+\int {dx_1\over x_1}\int dx_2\,\Phi^{\alpha\beta\gamma}_{F\,abc}(x_1,x_2)
{1\over x_2\mp i\epsilon}{1\over x_2-x_1\pm i\epsilon}S^{(\mp)[\pm]abc}_{R\mu\nu\lambda}(x_1p,x_2p)
\Bigr).
\label{formula_pp}
\eeq
It is natural that $\Phi_{\partial}^{[+]}$ appears in the formula and the dynamical part is expanded in terms of
all possible pole structures when the ISI contribution is additionally included.
Our work gives a systematic way to calculate the color factors $C^{f(d)(\pm)[\pm]}_{Lkl}$
and construct the hard part $S^{(\pm)[\pm]abc}_{L(R)\mu\nu\lambda}(x_1p,x_2p)$.
An unexpected feature is that the kinematical part contains 
$d^{abc}\Phi^{\alpha\beta\gamma}_{F\,abc}$. 
Because this term is associated with the coefficient $C^{d(-)[+]}_{Lkl}$, it can appear only in processes 
involving both ISI and FSI. As will be shown in the next subsection, this term is generally nonzero 
and is required to reproduce the known result for the $gq\to qg$-channel.

\subsection{Calculation of the hard cross sections}

In this section we verify that the formula (\ref{formula_pp}) reproduces the known result derived by the old pole technique for the quark fragmentation channel in $gq\to qg$.
The intrinsic part associated with $\Phi^{\alpha\beta}(x)$
does not contribute to naive $T$-odd observables such as the SSA.
We take the derivative of $S_{\mu\nu}(k)$ with respect to $k$ in the calculation of 
the kinematical part as
\beq
{\partial\over \partial k^{\lambda}}S_{kl\,\mu\nu}(k)\Bigr|_{k=xp}
={\partial\over \partial k^{\lambda}}H_{kl\,\mu\nu}(k)\Bigr|_{k=xp}
\delta\Bigl(\hat{s}+\hat{t}+\hat{u}\Bigr)
+{2(xp+x'p'-{P_h\over z})^{\lambda}\over 2p\cdot (x'p'-{P_h\over z})}H_{kl\,\mu\nu}(xp)
{d\over dx}\delta\Bigl(\hat{s}+\hat{t}+\hat{u}\Bigr).
\eeq
Then the kinematical part reads
\beq
&&\omega^{\mu}_{\ \alpha}\omega^{\nu}_{\ \beta}\omega^{\lambda}_{\ \gamma}
\sum_{k,l=1}^{3}\int {dx\over x^2}\delta\Bigl(\hat{s}+\hat{t}+\hat{u}\Bigr)\Bigl[
\Bigl(x{d\over dx}\Phi_{kl}^{\alpha\beta\gamma}(x)
-2\Phi_{kl}^{\alpha\beta\gamma}(x)\Bigr)
{2(x'p'-{P_h\over z})^{\lambda}\over \hat{u}}H_{kl\,\mu\nu}(xp)
\nonumber\\
&&+\Phi_{kl}^{\alpha\beta\gamma}(x)
\Bigl({\partial\over \partial k^{\lambda}}H_{kl\,\mu\nu}(k)\Bigr|_{k=xp}
+{2(x'p'-{P_h\over z})^{\lambda}\over \hat{u}}x{d\over dx}H_{kl\,\mu\nu}(xp)\Bigr)\Bigr],
\eeq
where
\beq
\Phi_{kl}^{\alpha\beta\gamma}(x)=C^{f(-)[+]}_{Lkl}\Phi^{[+]\alpha\beta\gamma}_{\partial}(x)
+\Bigl(C^{f(-)[-]}_{Lkl}+C^{f(+)[-]}_{Lkl}\Bigr)\Phi^{[-]\alpha\beta\gamma}_{\partial}(x)
-2\pi i C^{d(-)[+]}_{Lkl}d^{abc}\Phi^{\alpha\beta\gamma}_{F\,abc}(x,x).
\eeq
The parametrizations of $\Phi^{[\pm]\alpha\beta\gamma}_{\partial}(x)$
and $\Phi^{\alpha\beta\gamma}_{F\,abc}(x,x)$ are given in \cite{Koike:2019zxc},
\beq
\Phi^{[\pm]\alpha\beta\gamma}_{\partial}(x)
={M_N\over 2}g^{\alpha\beta}_{\perp}\epsilon^{pnS_{\perp}\gamma}G^{[\pm](1)}_T(x)
+{M_N\over 8}\Bigl(\epsilon^{pnS_{\perp}\{\alpha}g^{\beta\}\gamma}_{\perp}
+\epsilon^{pn\gamma\{\alpha}S^{\beta\}}_{\perp}\Bigr)\Delta H^{[\pm](1)}_{T}(x)+\cdots,
\eeq
\beq
\Phi^{\alpha\beta\gamma}_{F\,abc}(x_1,x_2)
={-if^{abc}\over N_c(N_c^2-1)}N^{\alpha\beta\gamma}(x_1,x_2)
+{N_cd^{abc}\over (N_c^2-4)(N_c^2-1)}O^{\alpha\beta\gamma}(x_1,x_2),
\eeq
\beq
N^{\alpha\beta\gamma}(x_1,x_2)&=&
2iM_N\Bigl[g_{\perp}^{\alpha\beta}\epsilon^{\gamma pnS_{\perp}}N(x_1,x_2)
-g_{\perp}^{\beta\gamma}\epsilon^{\alpha pnS_{\perp}}N(x_2,x_2-x_1)
-g_{\perp}^{\alpha\gamma}\epsilon^{\beta pnS_{\perp}}N(x_1,x_1-x_2)\Bigr]+\cdots,\hspace{5mm}
\\
O^{\alpha\beta\gamma}(x_1,x_2)
&=&2iM_N\Bigl[g_{\perp}^{\alpha\beta}\epsilon^{\gamma pnS_{\perp}}O(x_1,x_2)
+g_{\perp}^{\beta\gamma}\epsilon^{\alpha pnS_{\perp}}O(x_2,x_2-x_1)
+g_{\perp}^{\alpha\gamma}\epsilon^{\beta pnS_{\perp}}O(x_1,x_1-x_2)\Bigr]+\cdots.
\eeq
These dynamical functions satisfy the following symmetries.
\beq
O(x_1,x_2)&=&O(x_2,x_1),\hspace{5mm}O(x_1,x_2)=O(-x_1,-x_2),
\nonumber\\
N(x_1,x_2)&=&N(x_2,x_1),\hspace{5mm}N(x_1,x_2)=-N(-x_1,-x_2).
\label{symmetries}
\eeq
The kinematical functions have the following relations with the dynamical functions.
\beq
G_T^{[\pm](1)}(x)=\pm 4\pi(N(x,x)-N(x,0)),\hspace{5mm}\Delta H_T^{[\pm](1)}(x)=\mp 8\pi N(x,0),
\eeq
which allows us to express the cross section solely in terms of the dynamical functions.
The kinematical part takes the following form.
\beq
&&\omega^{\mu}_{\ \alpha}\omega^{\nu}_{\ \beta}\omega^{\lambda}_{\ \gamma}
\int{dx\over x^2}\Phi_{kl}^{\alpha\beta\gamma}(x)
{\partial\over \partial k^{\lambda}}S_{kl\,\mu\nu}(k)\Bigr|_{k=xp}
\nonumber\\
&=&M_N\int {dx\over x}\delta\Bigl(\hat{s}+\hat{t}+\hat{u}\Bigr)\Bigl[
\Bigl({d\over dx}G_T^{[\pm](1)}(x)-2{G_T^{[\pm](1)}(x)\over x}\Bigr)\hat{\sigma}^{[\pm]}_{dk1}
+\Bigl({d\over dx}\Delta H_T^{[\pm](1)}(x)-2{\Delta H_T^{[\pm](1)}(x)\over x}\Bigr)\hat{\sigma}^{[\pm]}_{dk2}
\nonumber\\
&&+\pi\Bigl({d\over dx}O(x,x)-2{O(x,x)\over x}\Bigr)\hat{\sigma}_{dko1}
+\pi\Bigl({d\over dx}O(x,0)-2{O(x,0)\over x}\Bigr)\hat{\sigma}_{dko2}
\nonumber\\
&&+{G_T^{[\pm](1)}(x)\over x}\hat{\sigma}^{[\pm]}_{ndk1}
+{\Delta H_T^{[\pm](1)}(x)\over x}\hat{\sigma}^{[\pm]}_{ndk2}
+\pi{O(x,x)\over x}\hat{\sigma}_{ndko1}+\pi{O(x,0)\over x}\hat{\sigma}_{ndko2}
\Bigr]
\nonumber\\
&=&\pi M_N\int {dx\over x}\delta\Bigl(\hat{s}+\hat{t}+\hat{u}\Bigr)\Bigl[
{d\over dx}O(x,x)\hat{\sigma}_{dko1}+
{d\over dx}N(x,x)\Bigl(4\hat{\sigma}^{[+]}_{dk1}-4\hat{\sigma}^{[-]}_{dk1}\Bigr)
+{d\over dx}O(x,0)\hat{\sigma}_{dko2}
\nonumber\\
&&+{d\over dx}N(x,0)\Bigl(-4\hat{\sigma}^{[+]}_{dk1}+4\hat{\sigma}^{[-]}_{dk1}
-8\hat{\sigma}^{[+]}_{dk2}+8\hat{\sigma}^{[-]}_{dk2}\Bigr)
+{O(x,x)\over x}\Bigl(\hat{\sigma}_{ndko1}-2\hat{\sigma}_{dko1}\Bigr)
\nonumber\\
&&+{N(x,x)\over x}\Bigl(-8\hat{\sigma}^{[+]}_{dk1}+8\hat{\sigma}^{[-]}_{dk1}
+4\hat{\sigma}^{[+]}_{ndk1}-4\hat{\sigma}^{[-]}_{ndk1}\Bigr)
+{O(x,0)\over x}\Bigl(\hat{\sigma}_{ndko2}-2\hat{\sigma}_{dko2}\Bigr)
\nonumber\\
&&
+{N(x,0)\over x}\Bigl(8\hat{\sigma}^{[+]}_{dk1}-8\hat{\sigma}^{[-]}_{dk1}
+16\hat{\sigma}^{[+]}_{dk2}-16\hat{\sigma}^{[-]}_{dk2}
-4\hat{\sigma}^{[+]}_{ndk1}+4\hat{\sigma}^{[-]}_{ndk1}
-8\hat{\sigma}^{[+]}_{ndk2}+8\hat{\sigma}^{[-]}_{ndk2}
\Bigr)\Bigr].
\label{kinematical}
\eeq
Each hard cross section is generally gauge-dependent. The results in the Feynman gauge are
listed in the appendix.
Note that some of the hard cross sections depend on the choice of the arbitrary vector $n$.
This $n$-dependence is canceled by combining them with the hard cross sections in the dynamical part. 
The hard parts of the dynamical part are given by
\beq
S_{L\mu\nu\lambda}^{(\mp)[\pm]abc}(x'p,xp)&=&
H_{L\mu\nu\lambda}^{(\mp)[\pm]abc}(x'p,xp)
\delta\Bigl(\hat{s}+\hat{t}+\hat{u}\Bigr),
\\
S_{R\mu\nu\lambda}^{(\mp)[\pm]abc}(xp,x'p)&=&
H_{R\mu\nu\lambda}^{(\mp)[\pm]abc}(xp,x'p)
\delta\Bigl(\hat{s}+\hat{t}+\hat{u}\Bigr),
\eeq
where we set $x_1=x'$ and $x_2=x$ for the $S_{L}$ part and $x_1=x$ and $x_2=x'$ for the $S_{R}$ part.
The identity
\beq
{x\over (x'\pm i\epsilon)(x-x'\pm i\epsilon)}={1\over x'\pm i\epsilon}+{1\over x-x'\pm i\epsilon},
\eeq
allows us to disentangle the two types of the poles.
Furthermore, changing the integration variable $x'$ to $x'\to x-x'$, the dynamical part is written solely in terms of
$1/(x-x'\pm i\epsilon)$ poles. The combination of $H_L$ and $H_R$ extracts the following pole contributions.
\beq
{1\over x-x'\mp i\epsilon}-{1\over x-x'\pm i\epsilon}&=&\pm 2\pi i\delta(x-x'),
\\
{1\over (x-x'\mp i\epsilon)^2}-{1\over (x-x'\pm i\epsilon)^2}&=&\pm 2\pi i
{\partial\over \partial x'}\delta(x-x').
\eeq
The dynamical part takes the form of
\beq
&&-{1\over 2}\omega^{\mu}_{\ \alpha}\omega^{\nu}_{\ \beta}\omega^{\lambda}_{\ \gamma}
\int {dx\over x}\int {dx'}
\Bigl(
\,\Phi^{\alpha\beta\gamma}_{F\,abc}(x',x)
{1\over x'\mp i\epsilon}{1\over x-x'\pm i\epsilon}S^{(\mp)[\pm]abc}_{L\mu\nu\lambda}(x'p,xp)
\nonumber\\
&&
+\Phi^{\alpha\beta\gamma}_{F\,abc}(x,x')
{1\over x'\mp i\epsilon}{1\over x'-x\pm i\epsilon}S^{(\mp)[\pm]abc}_{R\mu\nu\lambda}(xp,x'p)
\Bigr)
\nonumber\\
&=&\pi M_N\int {dx\over x}\delta\Bigl(\hat{s}+\hat{t}+\hat{u}\Bigr)\Bigl[
{d\over dx}O(x,x)\hat{\sigma}_{do1}+{d\over dx}N(x,x)\hat{\sigma}_{dn1}
+{d\over dx}O(x,0)\hat{\sigma}_{do2}+{d\over dx}N(x,0)\hat{\sigma}_{dn2}
\nonumber\\
&&+O(x,x)\hat{\sigma}_{o1}+N(x,x)\hat{\sigma}_{n1}
+O(x,0)\hat{\sigma}_{o2}+N(x,0)\hat{\sigma}_{n2}
\Bigr].
\label{dynamical}
\eeq
The results for the hard cross sections in the Feynman gauge are
listed in the appendix.
Combining (\ref{kinematical}) and (\ref{dynamical}), we obtain the final result,
\beq
P^0_{h}{d\Delta\sigma\over d^3\vec{P}_h}
&=&{\pi M_N\alpha_s\over S}\int {dx'\over x'}q(x')\int {dz\over z^2}D(z)\int {dx\over x}
\delta(\hat{s}+\hat{t}+\hat{u})\Bigl(\hat{t}x'\epsilon^{p'pnS_{\perp}}+{1\over z}\hat{s}\epsilon^{P_hpnS_{\perp}}\Bigr)
\nonumber\\
&&\times\Bigl[\Bigl({d\over dx}O(x,x)-{2O(x,x)\over x}+{d\over dx}O(x,0)-{2O(x,0)\over x}\Bigr)\hat{\sigma}_O
\nonumber\\
&&+\Bigl({d\over dx}N(x,x)-{2N(x,x)\over x}-{d\over dx}N(x,0)+{2N(x,0)\over x}\Bigr)\hat{\sigma}_N
\Bigr],
\eeq
where the hard cross sections are given by
\beq
\hat{\sigma}_O&=&\Bigl(-{4C_F\over N_c}{1\over \hat{s}\hat{t}}
+{2\over \hat{u}^2}\Bigr){(\hat{s}-\hat{t})\over \hat{u}}\Bigl({\hat{s}\over \hat{t}}
+{\hat{t}\over \hat{s}}\Bigr),
\\
\hat{\sigma}_N&=&\Bigl({4C_F\over N_c}{1\over \hat{s}\hat{t}}
-{3\over \hat{u}^2}\Bigr)\Bigl({\hat{s}\over \hat{t}}
+{\hat{t}\over \hat{s}}\Bigr).
\eeq
We find that the result is $n$-independent,
\beq
\Bigl(\hat{t}x'\epsilon^{p'pnS_{\perp}}+{1\over z}\hat{s}\epsilon^{P_hpnS_{\perp}}\Bigr)
=2{xx'\over z}\epsilon^{P_hpp'S_{\perp}},
\eeq
which follows from the identity
\beq
g^{\alpha\beta}\epsilon^{\mu\nu\rho\sigma}
=g^{\alpha\mu}\epsilon^{\beta\nu\rho\sigma}+g^{\alpha\nu}\epsilon^{\mu\beta\rho\sigma}
+g^{\alpha\rho}\epsilon^{\mu\nu\beta\sigma}+g^{\alpha\sigma}\epsilon^{\mu\nu\rho\beta}.
\eeq
This result is consistent with (9), (13) and (14) in \cite{Beppu:2013uda}
\footnote{There is a typo in (14) in \cite{Beppu:2013uda}. The overall sign in the right-hand side of the equation should be reversed.}.


\section{Summary}

In this work, we revisit the calculation of the three-gluon distribution contribution to the SSA in $pp$ collisions within the framework of the nonpole method, whose theoretical formulation has been substantially developed over the past several years.
Although the nonpole formalism has already been established for SIDIS where only FSI exists, its straightforward extension to processes including both ISI and FSI such as $pp$ collisions
has been hindered by an ambiguity in the sign of the $i\epsilon$ prescription appearing in the 
relations associated with the WTIs.
In this paper, we first reconsider the WTIs and show that, in $pp$ collisions, the general WTIs can be decomposed into minimal WTIs for the ISI and FSI subsets separately. 
We demonstrate that the sign of the $i\epsilon$ is uniquely determined in the relations for each subset, 
which allows us to extend the formalism established in SIDIS to $pp$ collisions.
Using these relations, we extract the twist-3 contribution for each of the four subsets and successfully construct gauge-invariant matrix elements. The resulting cross section formula is not simply 
the sum of the ISI and FSI contributions. In particular, we find the contribution
$C^{d(-)[+]}_{Lkl}d^{abc}\Phi^{\alpha\beta\gamma}_{F,abc}(x,x)$
that arises only through the coexistence of ISI and FSI.
In the nonpole approach, the cross section can generally be expressed in terms of three types of contributions, intrinsic, kinematical, and dynamical, as in the case of the Collins type contribution associated with twist-3 fragmentation functions. For the Sivers type contribution, we find that the intrinsic contribution vanishes,
the kinematical functions are replaced with the dynamical functions, and then the known result 
obtained with the old pole technique is successfully reproduced.
The formulation based on the nonpole method enables both Sivers type and Collins type contributions to be calculated within the same framework. Moreover, recent studies have shown that this approach, in which the kinematical contribution plays an essential intermediate role, can simplify NLO calculations. The method is therefore expected to be important also from a practical point of view.
Understanding the gluon Sivers effect is one of the major objectives of future experiments. We expect that the present work, which provides a comprehensive set of calculational techniques required within the collinear factorization framework, will serve as a theoretical basis for further developments in this direction.


\appendix

\section{Explicit forms of the hard parts}

The hard cross sections in (\ref{kinematical}) are given in the Feynman gauge by
\beq
\hat{\sigma}^{[+]}_{dk1}&=&\Bigl[{C_F\over N_c}{1\over \hat{u}}\Bigl({\hat{s}\over \hat{t}}
+{\hat{t}\over \hat{s}}\Bigr)-{\hat{s}^3+2\hat{s}^2\hat{t}+\hat{s}\hat{t}^2
+2\hat{t}^3\over 2\hat{t}\hat{u}^3}\Bigr]
\Bigl(x'\epsilon^{p'pnS_{\perp}}-{1\over z}\epsilon^{P_hpnS_{\perp}}\Bigr),
\nonumber\\
\hat{\sigma}^{[-]}_{dk1}&=&
\Bigl[{C_F\over N_c}{1\over \hat{u}}\Bigl({\hat{s}\over \hat{t}}
+{\hat{t}\over \hat{s}}\Bigr)+{\hat{s}^3-2\hat{s}^2\hat{t}+\hat{s}\hat{t}^2
-2\hat{t}^3\over 2\hat{t}\hat{u}^3}\Bigr]
\Bigl(x'\epsilon^{p'pnS_{\perp}}-{1\over z}\epsilon^{P_hpnS_{\perp}}\Bigr),
\nonumber\\
\hat{\sigma}^{[+]}_{dk2}&=&\Bigl[{C_F\over 2N_c}{1\over \hat{s}}\Bigl({\hat{s}\over \hat{t}}
+{\hat{t}\over \hat{s}}\Bigr)-{1\over 4}{(\hat{s}+2\hat{t})(\hat{s}^2+\hat{t}^2)
\over  \hat{s}\hat{t}\hat{u}^2}
\Bigr]x'\epsilon^{p'pnS_{\perp}},
\nonumber\\
\hat{\sigma}^{[-]}_{dk2}&=&\Bigl[{C_F\over 2N_c}{1\over \hat{s}}\Bigl({\hat{s}\over \hat{t}}
+{\hat{t}\over \hat{s}}\Bigr)+{1\over 4}{(\hat{s}-2\hat{t})(\hat{s}^2+\hat{t}^2)
\over  \hat{s}\hat{t}\hat{u}^2}
\Bigr]x'\epsilon^{p'pnS_{\perp}},
\nonumber\\
\hat{\sigma}_{dko1}&=&\Bigl[-{8C_F\over N_c}{1\over \hat{u}}\Bigl({\hat{s}\over \hat{t}}
+{\hat{t}\over \hat{s}}\Bigr)+4{\hat{s}^3+2\hat{s}^2\hat{t}+\hat{s}\hat{t}^2
+2\hat{t}^3\over \hat{t}\hat{u}^3}\Bigr]
\Bigl(x'\epsilon^{p'pnS_{\perp}}-{1\over z}\epsilon^{P_hpnS_{\perp}}\Bigr),
\nonumber\\
\hat{\sigma}_{dko2}&=&\Bigl[{8C_F\over N_c}{1\over \hat{s}\hat{u}}\Bigl({\hat{s}\over \hat{t}}
+{\hat{t}\over \hat{s}}\Bigr)-4{\hat{s}^3+2\hat{s}^2\hat{t}+\hat{s}\hat{t}^2
+2\hat{t}^3\over \hat{s}\hat{t}\hat{u}^3}\Bigr]
\Bigl(\hat{t}x'\epsilon^{p'pnS_{\perp}}+{1\over z}\hat{s}\epsilon^{P_hpnS_{\perp}}\Bigr),
\nonumber\\
\hat{\sigma}^{[+]}_{ndk1}&=&\Bigl[{C_F\over N_c}{1\over \hat{t}^2}
+{2\hat{s}^4+6\hat{s}^3\hat{t}+7\hat{s}^2\hat{t}^2+3\hat{t}^4\over 4\hat{s}\hat{t}^2\hat{u}^3}\Bigr]
\Bigl(\hat{t}x'\epsilon^{p'pnS_{\perp}}+{1\over z}\hat{s}\epsilon^{P_hpnS_{\perp}}\Bigr),
\nonumber\\
\hat{\sigma}^{[-]}_{ndk1}&=&\Bigl[{C_F\over N_c}{1\over \hat{t}^2}
-{2\hat{s}^4+4\hat{s}^3\hat{t}-7\hat{s}^2\hat{t}^2+6\hat{s}\hat{t}^3
-3\hat{t}^4\over 4\hat{s}\hat{t}^2\hat{u}^3}\Bigr]
\Bigl(\hat{t}x'\epsilon^{p'pnS_{\perp}}+{1\over z}\hat{s}\epsilon^{P_hpnS_{\perp}}\Bigr),
\nonumber\\
\hat{\sigma}^{[+]}_{ndk2}&=&\Bigl[-{C_F\over N_c}{1\over \hat{t}}
-{2\hat{s}^4+6\hat{s}^3\hat{t}+5\hat{s}^2\hat{t}^2+\hat{t}^4\over 4\hat{s}\hat{t}\hat{u}^3}
\Bigr]x'\epsilon^{p'pnS_{\perp}}
+\Bigl[{C_F\over N_c}{1\over \hat{s}}
+{\hat{s}^2+6\hat{s}\hat{t}+3\hat{t}^2\over 4\hat{u}^3}
\Bigr]{1\over z}\epsilon^{P_hpnS_{\perp}},
\nonumber\\
\hat{\sigma}^{[-]}_{ndk2}&=&\Bigl[-{C_F\over N_c}{1\over \hat{t}}
+{2\hat{s}^4-5\hat{s}^2\hat{t}^2+2\hat{s}\hat{t}^3-\hat{t}^4\over 4\hat{s}\hat{t}\hat{u}^3}
\Bigr]x'\epsilon^{p'pnS_{\perp}}
+\Bigl[{C_F\over N_c}{1\over \hat{s}}
+{2\hat{s}^3-3\hat{s}^2\hat{t}+4\hat{s}\hat{t}^2+3\hat{t}^3\over 4\hat{t}\hat{u}^3}
\Bigr]{1\over z}\epsilon^{P_hpnS_{\perp}},
\nonumber\\
\hat{\sigma}_{ndko1}&=&-\Bigl[{8C_F\over N_c}{1\over \hat{t}^2}
+4{2\hat{s}^4+6\hat{s}^3\hat{t}+7\hat{s}^2\hat{t}^2+3\hat{t}^4\over 2\hat{s}\hat{t}^2\hat{u}^3}\Bigr]
\Bigl(\hat{t}x'\epsilon^{p'pnS_{\perp}}+{1\over z}\hat{s}\epsilon^{P_hpnS_{\perp}}\Bigr),
\nonumber
\eeq
\beq
\hat{\sigma}_{ndko2}&=&\Bigl[{8C_F\over N_c}{1\over \hat{t}}
+4{2\hat{s}^4+6\hat{s}^3\hat{t}+3\hat{s}^2\hat{t}^2-\hat{t}^4
\over 2\hat{s}\hat{t}\hat{u}^3}\Bigr]
x'\epsilon^{p'pnS_{\perp}}
\nonumber\\
&&-\Bigl[{8C_F\over N_c}{\hat{s}^2+2\hat{t}^2\over \hat{s}\hat{t}^2}
+4{2\hat{s}^4+6\hat{s}^3\hat{t}+9\hat{s}^2\hat{t}^2+12\hat{s}\hat{t}^3
+9\hat{t}^4\over 2\hat{t}^2\hat{u}^3}
\Bigr]{1\over z}\epsilon^{P_hpnS_{\perp}}.
\eeq
The hard cross sections in (\ref{dynamical}) are given in the Feynman gauge by
\beq
\hat{\sigma}_{do1}&=&-{2C_F\over N_c}{1\over \hat{s}\hat{t}}
\Bigl({\hat{s}\over \hat{t}}+{\hat{t}\over \hat{s}}\Bigr)
\Bigl(\hat{t}x'\epsilon^{p'pnS_{\perp}}-{1\over z}\hat{s}\epsilon^{P_hpnS_{\perp}}\Bigr)
+{2\over \hat{u}^2}\Bigl({\hat{s}\over \hat{t}}+{\hat{t}\over \hat{s}}\Bigr)
\Bigl((\hat{s}-\hat{u})x'\epsilon^{p'pnS_{\perp}}-{3\over z}\hat{s}\epsilon^{P_hpnS_{\perp}}\Bigr),
\nonumber\\
\hat{\sigma}_{do2}&=&\Bigl[{4C_F\over N_c}{\hat{s}^2+\hat{t}^2\over \hat{s}^2\hat{t}^2}
-{6\over \hat{u}^2}\Bigl({\hat{s}\over \hat{t}}+{\hat{t}\over \hat{s}}\Bigr)
\Bigr]\Bigl(\hat{t}x'\epsilon^{p'pnS_{\perp}}+{1\over z}\hat{s}\epsilon^{P_hpnS_{\perp}}\Bigr),
\nonumber\\
\hat{\sigma}_{dn1}&=&{4C_F\over N_c}{1\over \hat{s}\hat{t}}
\Bigl({\hat{s}\over \hat{t}}+{\hat{t}\over \hat{s}}\Bigr)
\Bigl(\hat{t}x'\epsilon^{p'pnS_{\perp}}+{1\over z}\hat{s}\epsilon^{P_hpnS_{\perp}}\Bigr)
+{2\over \hat{u}^3}\Bigl({\hat{s}\over \hat{t}}+{\hat{t}\over \hat{s}}\Bigr)
\Bigl((2\hat{s}^2+3\hat{s}\hat{t}+3\hat{t}^2)x'\epsilon^{p'pnS_{\perp}}+
{1\over z}(\hat{s}^2+3\hat{s}\hat{t})\epsilon^{P_hpnS_{\perp}}\Bigr),
\nonumber\\
\hat{\sigma}_{dn2}&=&-\Bigl[{4C_F\over N_c}{\hat{s}^2+\hat{t}^2\over \hat{s}^2\hat{t}^2}
+{2}{\hat{s}^3+3\hat{s}^2\hat{t}+\hat{s}\hat{t}^2+3\hat{t}^3\over \hat{s}\hat{t}\hat{u}^3}\Bigr]
\Bigl(\hat{t}x'\epsilon^{p'pnS_{\perp}}+{1\over z}\hat{s}\epsilon^{P_hpnS_{\perp}}\Bigr),
\nonumber\\
\hat{\sigma}_{o1}&=&\Bigl[{8C_F\over N_c}{2\hat{s}^2+\hat{t}^2\over \hat{s}^2\hat{t}}
+2{6\hat{s}^4+12\hat{s}^3\hat{t}+13\hat{s}^2\hat{t}^2+6\hat{s}\hat{t}^3
+5\hat{t}^4\over \hat{s}\hat{t}\hat{u}^3}
\Bigr]x'\epsilon^{p'pnS_{\perp}}
\nonumber\\
&&+\Bigl[-{8C_F\over N_c}{1\over \hat{s}}+2{2\hat{s}^4+\hat{s}^2\hat{t}^2
-6\hat{s}\hat{t}^3-3\hat{t}^4\over \hat{t}^2\hat{u}^3}\Bigr]
{1\over z}\epsilon^{P_hpnS_{\perp}},
\nonumber\\
\hat{\sigma}_{o2}&=&-\Bigl[{8C_F\over N_c}{2\hat{s}^2+\hat{t}^2\over \hat{s}^2\hat{t}}
+2{2\hat{s}^4+12\hat{s}^3\hat{t}+9\hat{s}^2\hat{t}^2+6\hat{s}\hat{t}^3
+5\hat{t}^4\over \hat{s}\hat{t}\hat{u}^3}
\Bigr]x'\epsilon^{p'pnS_{\perp}}
\nonumber\\
&&+\Bigl[{8C_F\over N_c}{1\over \hat{s}}+2{2\hat{s}^4+3\hat{s}^2\hat{t}^2
+6\hat{s}\hat{t}^3+3\hat{t}^4\over \hat{t}^2\hat{u}^3}\Bigr]
{1\over z}\epsilon^{P_hpnS_{\perp}},
\nonumber\\
\hat{\sigma}_{n1}&=&-\Bigl[{8C_F\over N_c}{\hat{s}^2+\hat{t}^2\over \hat{s}^2\hat{t}}
+2{6\hat{s}^4+11\hat{s}^3\hat{t}+10\hat{s}^2\hat{t}^2+9\hat{s}\hat{t}^3
+6\hat{t}^4\over \hat{s}\hat{t}\hat{u}^3}\Bigr]x'\epsilon^{p'pnS_{\perp}}
\nonumber\\
&&-\Bigl[{8C_F\over N_c}{\hat{s}^2+\hat{t}^2\over \hat{s}\hat{t}^2}
+2{2\hat{s}^4+7\hat{s}^3\hat{t}+6\hat{s}^2\hat{t}^2
+5\hat{s}\hat{t}^3+6\hat{t}^4\over \hat{t}^2\hat{u}^3}\Bigr]
{1\over z}\epsilon^{P_hpnS_{\perp}},
\nonumber\\
\hat{\sigma}_{n2}&=&\Bigl[{8C_F\over N_c}{\hat{s}^2+\hat{t}^2\over \hat{s}^2\hat{t}}
+2{-2\hat{s}^4+\hat{s}^3\hat{t}+6\hat{s}^2\hat{t}^2+3\hat{s}\hat{t}^3
+6\hat{t}^4\over \hat{s}\hat{t}\hat{u}^3}\Bigr]x'\epsilon^{p'pnS_{\perp}}
\nonumber\\
&&+\Bigl[{8C_F\over N_c}{\hat{s}^2+\hat{t}^2\over \hat{s}\hat{t}^2}
+2{2\hat{s}^4+5\hat{s}^3\hat{t}+10\hat{s}^2\hat{t}^2
+7\hat{s}\hat{t}^3+6\hat{t}^4\over \hat{t}^2\hat{u}^3}\Bigr]
{1\over z}\epsilon^{P_hpnS_{\perp}}.
\eeq


\section*{Acknowledgements}

This work is supported by Polish National Science Center Grant No. UMO-2023/49/B/ST2/03665.

\end{document}